\documentclass[aps,prb,superscriptaddress,twocolumn,showpacs,amsmath,floatfix,citeautoscript]{revtex4-2}

\usepackage{siunitx}
\usepackage{graphicx}
\usepackage{float}
\usepackage{subfigure}
\usepackage{dcolumn}
\usepackage{bm}
\usepackage{color}
\usepackage{latexsym,bm}
\usepackage[normalem]{ulem}
\usepackage{multirow}
\usepackage{appendix}
\usepackage{booktabs}
\usepackage{array}    
\usepackage{threeparttable}     
\usepackage[colorlinks=true, citecolor=black, urlcolor=black, linkcolor=black]{hyperref}     

\newcommand{\bfr}{ {\bf r}} 
\newcommand{\bfrp}{ {\bf r'}} 
\newcommand{\bfR}{ {\bf R}} 
\newcommand{\bfk}{ {\bf k}} 

\begin{document}

\hyphenpenalty=5000

\tolerance=1000

\title{DFT+$U$ within the framework of linear combination of numerical atomic orbitals}

\author{Xin Qu}
\affiliation{Rocket Force University of Engineering, Xi'an, Shaanxi 710025, China}
\affiliation{CAS Key Laboratory of Quantum Information, University of Science and Technology of
China, Hefei, Anhui 230026, China}

\author{Peng Xu}
\affiliation{Rocket Force University of Engineering, Xi'an, Shaanxi 710025, China}
\author{Hong Jiang}
\email{jianghchem@pku.edu.cn}
\affiliation{Beijing National Laboratory for Molecular Sciences, College of Chemistry and Molecular Engineering, Peking University, 100871 Beijing, China}
\author{Lixin He}
\email{helx@ustc.edu.cn}
\affiliation{CAS Key Laboratory of Quantum Information, University of Science and Technology of
China, Hefei, Anhui 230026, China}
\author{Xinguo Ren}
\email{renxg@iphy.ac.cn}
\affiliation{Institute of Physics, Chinese Academy of Sciences,  Beijing 100190, China}
\affiliation{Songshan Lake Materials Laboratory, Dongguan 523808, Guangdong, China}

\date{\today }


\begin{abstract}
\par We present a formulation and implementation of the DFT+\textit{U} method within the framework of linear combination of numerical atomic orbitals (NAO). Our implementation not only enables single-point total energy and electronic-structure calculations but also provides access to atomic forces and stresses, hence allowing for full structure relaxations of periodic systems. 
Furthermore, our implementation allows one to deal with non-collinear spin texture, with 
the spin-orbit coupling (SOC) effect treated self-consistently. 
The key aspect behind our implementation is a suitable definition of
the correlated subspace when multiple atomic orbitals with the same angular momentum are used, and this is addressed via the ``Mullken charge projector"
constructed in terms of the first (most localized) atomic orbital within the $d/f$ angular momentum channel.
The important Hubbard $U$ and Hund $J$ parameters can be estimated from a screened Coulomb potential of the Yukawa type, 
with the screening parameter either chosen semi-empirically or determined from the Thomas-Fermi screening model. 
Benchmark calculations are performed for four late transition metal monoxide bulk systems, i.e., MnO, FeO, CoO, and NiO, and for the 5$d$-electron compounds 
IrO$_2$. For the former type of systems, we check the performance of our DFT+$U$ implementation for calculating band gaps, magnetic moments, electronic 
band structures, as well as forces and stresses; for the latter, the efficacy of our DFT+$U$+SOC implementation is assessed. 
Systematic comparisons with available experimental results, and especially with the results from 
other implementation schemes are carried out,
which demonstrate the validity of our NAO-based DFT+$U$ formalism and implementation. 
\end{abstract}
\maketitle

\date{\today}

\section{Introduction}
\par The density functional theory (DFT) developed by Hohenberg, Kohn and Sham \cite{Hohenberg1964, Kohn1965} uses the charge density as the basic variable to determine the ground state of interacting many-particle systems, which allows one to simulate systems of sizes that are prohibitively expensive for wave function based methods \cite{Kohn1999}. Within DFT, suitable approximations can be designed to achieve excellent balance between accuracy and efficiency, and
as such DFT has been widely used for electronic structure calculations for a large variety of physical and chemical problems. However, within the popular local (spin-) density approximation (L(S)DA) or generalized gradient approximations (GGAs), DFT encounters failures in several known situations -- in particular for strongly correlated materials, usually characterized by partially filled \textit{d}/\textit{f} electron states. 
These include transition metals (TM) and their oxides, rare-earth compounds, and lanthanides, to name a few, where L(S)DA/GGAs typically 
yield quantitatively or even qualitatively wrong results. 

\par The failure of L(S)DA or GGAs in strongly correlated materials can be traced back to the large delocalization error \cite{Cohen2008,Li2018,Su2018,Su2020}, more often termed as many-body self-interaction errors \cite{Perdew1982}, and the static correlation error \cite{Cohen2008, Anisimov2010,Himmetoglu2014} associated with these functionals. Many approaches have been proposed to address these deficiencies, such as self-interaction corrected DFT \cite{Perdew1981},  hybrid functionals \cite{Becke1993, Heyd2003},  the localized orbital scaling correction 
\cite{Li2018, Su2020}, fractional spin correction \cite{Su2018}, and so on.
Among these, the most popular approaches in solid-state physics are the combination of L(S)DA and GGAs with the non-perturbative many-body technique -- 
dynamic mean-field theory (DMFT)  \cite{Metzner1989, Georges1996, Anisimov1997_DMFT, Lichtenstein1998, Kotliar2006, Held2007} and with a simpler mean-field-type correction based on the Hubbard model \cite{Hubbard1963, Gutzwiller1963, Kanamori1963}. The latter approach, commonly known as DFT+\textit{U} \cite{Anisimov1991, Anisimov1993, Anisimov1997_LDAU}, inherits the efficiency of L(S)DA/GGA, but gains the strength of the Hubbard model in describing the physics of strongly correlated systems. Owing to its success in describing certain non-trivial properties arising from strong correlation, e.g. band-gap opening in transition metal oxides (TMO) and/or rare-earth compounds at a similar cost as L(S)DA/GGAs, DFT+\textit{U} has become one of the most widely used first-principles approaches for strongly correlated (especially insulating) systems.

\par Being enormously successful in dealing with open-shell systems with partially filled $d$/$f$ states, DFT+\textit{U} has become
a standard module available in many DFT code packages, based on various basis-set frameworks. For instance, it has been implemented 
within the full-potential all-electron linearized augmented plane-wave (LAPW) framework, as exemplified by the WIEN2k \cite{Shick1999} and ELK \cite{ELK} codes,
and within the projector-augmented-wave (PAW) or norm-conserving pseudopotential based plane-wave framework, with VASP \cite{Bengone2000}, 
Abinit \cite{Amadon2008}, and  QUANTUM ESPRESSO \cite{Giannozzi2009, Cococcioni2005} as prominent examples. In recent years, the linear combination
of numerical atomic orbitals (NAOs) has emerged as a versatile basis set framework for implementing both conventional local and semi-local density functional approximations \cite{Delley2000,Koepernik1999,Soler2002a,Ozaki/etal:2008,Volker2009,Li2016a},
hybrid functionals \cite{Ren2012, Levchenko2015, Lin2020, Lin2021}, and many-body perturbation
theories \cite{Ren2012,Tahir2019,Ren2021}.
 Compared to other basis set choices, NAOs are considered to be advantageous for simulating large-scale systems, due to their compact size and strict locality
 in real space. Thus the implementation of the DFT+\textit{U} method within the NAO framework is of great interest. Similar to other numerical schemes, 
 the key aspect for the DFT+\textit{U} implementation is to define a suitable projector, which maps the full-orbital space where
 the first-principles DFT calculations are performed to a local, correlated subspace where the local orbitals behind the (generalized) Hubbard model 
 are defined. The NAO-based DFT+$U$ implementations have been reported for  SIESTA \cite{Soler2002a, SIESTA}, OpenMX \cite{Han2006}, and
 FHI-aims codes \cite{Kick2019}, where the emphasis has been placed on the choice of the projector function, as well as their influence on the suitable $U$ values and
 the obtained results. Experiences strongly suggest that the DFT+$U$ cannot be taken as a black-box method and a proper use of this approach 
 requires a good understanding of the underlying technical aspects of a given implementation. 
 
 \par In this work, we report yet another implementation of the DFT+$U$ method within ABACUS \cite{Li2016a}, which is a first-principles code package 
 based on the norm-conserving pseudopotentials and NAO basis functions. ABACUS allows one to use plane-wave basis functions as well, but 
 our current DFT+$U$ implementation is based on the NAO basis set framework, and hence shares similarities with the OpenMX \cite{Han2006}
 and FHI-aims implementations \cite{Kick2019}. However, due to the different strategies adopted for basis set generation,
 the resultant NAOs differ in shapes and spatial extent, which further affects the projector and other implementation details.
 Our implementation not only supports usual self-consistent DFT+$U$ electronic structure calculations but also allows for
 force and stress computations, thus enabling full structural relaxations. Furthermore, non-collinear spin configurations and SOC effects can be
 treated within our implementation, which is instrumental for systems containing heavy elements. Last but not least, we have made attempts 
 to compute the Hubbard $U$ value on the fly from a Yukawa-type screened Coulomb potential. The performance of such a
 scheme for determining the $U$ value will be examined. Considering all these aspects, we believe that 
 a concise description of the formulation and numerical details of our implementation
 should be not only useful for the users and developers of the ABACUS code, but also of general interest to the electronic-structure
 community using the NAO basis sets.

\par The paper is organized as follows. In Sec.~\ref{sec:formalism} we present the detailed formulation behind our implementation, including
the rotationally invariant DFT+$U$ energy functional, our choice of the  local projector, the force and stress evaluations, 
as well as
the incorporation of SOC in the NAO-based DFT+$U$ formalism and the determination of the $U$, $J$ values from the Yukawa screened Coulomb potential.
Sec.~\ref{sec:details} presents the computational details in this paper.
In Sec~\ref{sec:results}, the efficacy of our formalism and the validity of our implementation will be examined. This is
done by comparing the results of our implementation to those of experiments and particularly of other well-tested codes.
Finally, we conclude this work in Sec.~\ref{sec:summary}.

\section{Formalism}
\label{sec:formalism}
\subsection{General DFT+\textit{U} functional}
\label{sec:general_formalism}
\par The basic idea of DFT+\textit{U} is to treat strongly correlated \textit{d}/\textit{f} electrons in terms of multi-orbital Hubbard model at the level of
static Hartree-Fock mean-field theory, whereas all the rest electrons are described at the level of conventional density functional approximations (DFAs) like L(S)DA and GGA. Since the $U$ correction term in DFT+\textit{U} corresponds to a Hartree-Fock approximation of the multi-orbital Hubbard model, the unphysical self-interactions among strongly correlated \textit{d}/\textit{f} electrons present in the Hartree energy 
are canceled out. This
is considered as the main reason responsible for the success of DFT+\textit{U} in producing more reliable insulating gaps, magnetic moments, 
and other properties for
TMOs \cite{Anisimov1997_LDAU}. The many-body complexity is encoded into the screened Coulomb interaction among strongly correlated \textit{d}/\textit{f} electrons, usually parameterized in terms of the Hubbard \textit{U} for describing the direct Coulomb interaction and the Hund \textit{J} for describing the exchange interaction. 

\par The formulation of the DFT+\textit{U} approach begins with the following energy functional
 \begin{equation} \label{eq:A1}
 	E_{\text{DFT}+U} = E_{\text{DFA}} + E_{U} - E_{\text{dc}} ~,
 \end{equation}
whereby $E_{\text{DFA}}$ is the energy of density functional approximations at the level of L(S)DA or GGA, and $E_U$ is the Coulomb interaction energy due to strongly correlated electrons given by the Hartree-Fock approximation to multi-orbital Hubbard model. The double counting term $E_{\text{dc}}$ is subtracted here to discount the Coulomb interaction energy
that is already included in DFAs at an average level.  

\par Now we discuss the last two terms on the right-hand side of Eq.~(\ref{eq:A1}). Using the second-quantization language, the full electron-electron interaction term can be written as
\begin{equation}\label{eq:A2}
	\begin{aligned}
		\hat{H}_{\text{Hub}} =& \frac{1}{2}\sum_{\{m\}} \sum_{\sigma \sigma^\prime} \langle m  m^\prime  \left| v_{sc} \right|  m^{\prime \prime} m^{\prime \prime \prime}\rangle \\
		& \times \hat{c}^{\sigma \dagger}_m \hat{c}^{\sigma^\prime \dagger}_{m^\prime} \hat{c}_{m^{\prime \prime \prime}}^{\sigma^\prime} \hat{c}_{m^{\prime \prime}}^{\sigma}
	\end{aligned} ~,
\end{equation}
where the Coulomb interaction matrix elements are
\begin{equation}\label{eq:A3}
	\begin{aligned}
	&\langle m  m^\prime  \left| v_{sc} \right|  m^{\prime \prime} m^{\prime \prime \prime}  \rangle\\
	=& \int d\textbf{r} \int d\textbf{r}^\prime \varphi_m^* \left( \textbf{r} \right) \varphi_{m^\prime}^* \left( \textbf{r}^\prime \right) v_{sc} \left( \textbf{r}, \textbf{r}^\prime\right) \varphi_{m^{\prime \prime}} \left( \textbf{r} \right) \varphi_{m^{\prime \prime \prime}} \left( \textbf{r}^\prime \right) ~.
\end{aligned} 
\end{equation}
 In Eqs.~(\ref{eq:A2}) and (\ref{eq:A3}), $v_{sc}(\bfr,\bfrp)$ is the (statically) screened Coulomb potential, $\{m\}$ the local orbital indices for \textit{d} or \textit{f} subshell, $\sigma$ the spin index and $\hat{c}_m^{\sigma\dagger}$ and $\hat{c}_m^{\sigma}$ the creation and annihilation operators associated with the local correlated orbitals. 

\par For simplicity, we neglect the spin-orbit coupling (SOC) effect at this stage. The generalized DFT+$U$ formalism 
that incorporates the SOC effect will be presented in Sec.~\ref{sec:formalism_soc}. 
In this case, each Kohn-Sham (KS) spin orbital is a product of a spatial function $\psi_{n \textbf{k}}\left(\textbf{r}\right)$ and a spin function. 
The ground state of the KS system is a Slater determinant formed by occupied KS spin-orbitals \cite{Martin2004}. The Hartree-Fock approximation to the Hubbard Hamiltonian Eq.~(\ref{eq:A2}) can be obtained by evaluating its expectation value within the
KS ground state $| 0 \rangle$, yielding the energy contribution as
\begin{equation}\label{eq:A5}
	\begin{aligned}
		E_{U}=& \langle 0| \hat{H}_{\text{Hub}} | 0 \rangle \\
		=& \frac{1}{2} \sum_{\{m\}, \sigma}\left\{\left(\left\langle m m^{\prime}\left|v_{s c}\right| m^{\prime \prime}  m^{\prime \prime \prime} \right\rangle\right.\right.\\
		&\left.-\left\langle m m^{\prime} \left|v_{\mathrm{sc}}\right| m^{\prime \prime \prime} m^{\prime \prime}  \rangle\right) \right. n_{m^{\prime \prime} m}^{\sigma} n_{m^{\prime \prime \prime} m^{\prime}}^{\sigma} \\
		&\left.+\left\langle m  m^{\prime} \left|v_{\mathrm{sc}}\right| m^{\prime \prime} m^{\prime \prime \prime} \right\rangle n_{m^{\prime \prime} m}^{\sigma} n_{m^{\prime \prime \prime} m^{\prime}}^{-\sigma}\right\}\, .
	\end{aligned}
\end{equation}
In Eq.~(\ref{eq:A5}), $n_{mm'}^\sigma$  is the local occupation matrix given by
\begin{equation}\label{eq:A6}
 \begin{aligned}
	n_{m m^\prime}^\sigma & = \langle 0 | \hat{c}^{\sigma \dagger}_{m'} \hat{c}^{\sigma}_{m} | 0 \rangle \\
	& =\frac{1}{N_{\textbf{k}}} \sum_{n \textbf{k}} f_{n  \textbf{k}}^\sigma \langle \psi_{n  \textbf{k}}^\sigma \sigma |m \sigma \rangle \langle m^\prime \sigma | \psi_{n \textbf{k}}^\sigma \sigma \rangle \, 
	 \end{aligned} 
\end{equation}
with  $f_{n  \textbf{k}}^\sigma$  being the occupation number of KS orbitals $| \psi_{n \textbf{k}}^\sigma \sigma \rangle$. Here $| \psi_{n \textbf{k}}^\sigma \sigma \rangle = |\psi_{n  \textbf{k}}^\sigma\rangle |\sigma\rangle$ is a product of the spatial wavevector $|\psi_{n  \textbf{k}}^\sigma\rangle $, which is the Kohn-Sham wavefunction of the $\sigma$-spin component in the position space, and the spin function $|\sigma\rangle$. Similarly the local spin orbital
$|m\sigma\rangle = |\phi_m\rangle |\sigma\rangle$.  
Furthermore, $N_{\textbf{k}}$ is the number of $\bfk$ points in the Brillouin zone (BZ), which equals the number of unit cells in the Born-Von-K\'{a}rm\'{e}n (BvK) supercell under the periodic boundary condition. 
Equation~(\ref{eq:A5}) is the well known rotationally invariant form of  DFT+\textit{U} firstly proposed by  Lichtenstein \textit{et al.} \cite{Lichtenstein1995} in 1995, whereby the local occupation number matrix $n_{m m^\prime}^\sigma$ is the key quantity.  


\par Since the local occupation matrix is symmetric, one can always introduce an unitary transformation to diagonalize it and arrives at
\begin{equation}\label{eq:A7}
	E_{U}= \frac{1}{2} U \sum_{m m^\prime, \sigma} \textbf{n}_{m}^{\sigma} \textbf{n}_{m^{\prime}}^{-\sigma} + \frac{1}{2} \left(U-J\right) \sum_{m\neq m^\prime, \sigma} \textbf{n}_{m}^{\sigma} \textbf{n}_{m^{\prime}}^{\sigma} .
\end{equation}
Here we use bold $\textbf{n}$ to denote the vector comprising the eigenvalues of the local occupation matrix. Furthermore, $U \equiv \langle m m^\prime\left| v_{sc} \right|  m m^{\prime} \rangle $ and $J \equiv \langle m m^\prime\left| v_{sc} \right| m^{\prime} m\rangle$ are the direct Coulomb and exchange 
integrals of the electrons in the correlated subspace, respectively, which are assumed to be isotropic, i.e., independent of the magnetic quantum number $m$. Note that the self-interaction is absent both in Eq.~(\ref{eq:A5}) and (\ref{eq:A7}). Theoretically, $\langle m m^\prime\left| v_{sc} \right|  m m^{\prime} \rangle$ can be evaluated through Slater integrals, but the detailed from of the screened Coulomb interaction $v_{\text{sc}}$ remains unknown, and therefore
in practical calculations $U$ and $J$ are most commonly treated as adjustable parameters
or obtained via pragmatic schemes like constrained DFT \cite{Dederichs1984,Norman1986,Gunnarsson1989,Anisimov1991}, constrained random-phase approximation (RPA) \cite{Aryasetiawan2004,Miyake2008,Miyake2009,Sakuma2013}, or linear-response approach \cite{Cococcioni2005}. 

\par The double-counting term $E_\text{dc}$ in Eq.~(\ref{eq:A1}) is an important portion of the DFT+$U$ theory and needs to be properly treated. 
Unfortunately, there are uncertainties for a rigorous definition of this term. This difficulty arises from by the fact that local/semi-local DFAs are 
not orbital-resolved theories, and contributions from individual orbitals cannot be separated from one another. By now there are two main double counting schemes used in practical DFT+\textit{U} calculations. One is the so-called
``around mean field (AMF)" scheme and another is the ``fully localized limit (FLL)" scheme \cite{Anisimov1991, Anisimov1993, Ylvisaker2009a}. Both schemes are
physically motivated. It's generally accepted that the former gives a better description of metallic systems while the latter one is more suitable for insulating systems \cite{Himmetoglu2014}. The FLL double counting term is given by
\begin{equation}\label{Eq:A8}
	E_{\text{dc}}= \frac{1}{2}U N \left( N-1\right)- \frac{1}{2} J \sum_{\sigma} N^{\sigma} \left( N^\sigma -1\right) ,
\end{equation}
where $N=\sum_{\sigma}N^\sigma$, and $N^\sigma$ is the total number of correlated \textit{d} or \textit{f} electrons of spin $\sigma$. The FLL double counting term 
can be derived by assuming integer occupations of correlated \textit{d} or \textit{f} electrons in the atomic limit. In our implementation, the FLL scheme is used.

\par Subtracting the double counting term $E_{\text{dc}}$ from $E_U$ and making some simple derivation, the DFT+\textit{U} energy correction can be explicitly expressed as 
\begin{equation}\label{eq:A9}
	\Delta E_{\text{DFT+}U} = E_{U} - E_{\text{dc}}= \frac{1}{2} \left( U - J \right) \sum_{m \sigma} \left( \textbf{n}_m^{\sigma} - \textbf{n}_m^{\sigma} \textbf{n}_m^{\sigma} \right) ~.
\end{equation}
Since the trace of an arbitrary matrix remains unchanged after the unitary transform,  $\Delta E_{\text{DFT+}U}$ can also be rewritten as
\begin{equation}\label{eq:A10}
\begin{aligned}
 \Delta E_{\text{DFT+}U} 
 =\frac{1}{2} \left( U - J \right) \sum_{\sigma} \left[ \sum_{m} n_{m m}^{\sigma} - \sum_{m m^\prime} n_{m m^\prime}^{\sigma} n_{m^\prime m}^{\sigma} \right] .
\end{aligned} 
\end{equation}
The above energy correction functionals, Eq.~(\ref{eq:A9}) and (\ref{eq:A10}), are the simplified form of the rotationally invariant scheme proposed by Dudarev \textit{et al.} \cite{Dudarev1998a},
while still retaining the rotational invariance as the energy correction stays unchanged under unitary transformations of the given set of correlated orbitals. Within this functional, the DFT+\textit{U} total energy reduces to standard L(S)DA/GGA in the case of empty or full (0 or 1) occupation of local orbitals. 

\subsection{Mulliken charge projector}
\label{sec:mulliken}

\par The DFT+$U$ formalism presented in Sec.~\ref{sec:general_formalism} only applies to the  single-site case, i.e., only one correlated atom in the cell. To deal with
the multi-site cases, it is necessary to introduce an extra correlated atomic index $I$ to label the local occupation matrix and the parameters \textit{U} and \textit{J}. Furthermore,
in Sec.~\ref{sec:general_formalism} it is implicitly assumed that the local correlated orbitals $\{|m\rangle\}$ are orthonormal to each other, and these are not necessarily satisfied for practically chosen local orbitals. Considering these complexities, it is convenient to introduce a local projection
operator $P_{mm'}^\sigma$, called ``projector", such that the spin-dependent local occupation matrix is given by
\begin{equation}
	n_{I,mm'}^\sigma = \frac{1}{N_{\textbf{k}}} \sum_{n \textbf{k}} f_{n  \textbf{k}}^\sigma
\langle \psi_{n  \textbf{k}}^\sigma \sigma| \hat{P}_{I,m m^\prime}^{\sigma} | \psi_{n \textbf{k}}^\sigma \sigma \rangle  \, .
\label{eq:projector2occupa}
\end{equation}
where $I$ denotes a correlated atom to which the $U$ correction needs to be applied.
The key issue in the implementation of DFT+\textit{U} is to construct such a projector that maps the full Kohn-Sham orbital space into the correlated subspace. The choice of the projector depends on the underlying computational frameworks, ranging from the linear muffin-tin orbital method \cite{Anisimov1997_LDAU}, the LAPW method \cite{Shick1999}, to the PAW method \cite{Amadon2008,Bengone2000} and
pseudopotential-based plane-wave method \cite{Cococcioni2005}. Within the NAO-based framework, the most straightforward way
is to utilize the local $d/f$-type atomic-orbital basis functions to construct the projector.  However, in practical 
calculations the NAOs centering on neighboring atoms have finite overlaps, i.e., they are non-orthogonal to each other.
This non-orthogonality has to  be taken
into account when defining a suitable projector. In this regard, we follow the previous work of Han \textit{et al}. \cite{Han2006} 
where the so-called ``Mulliken charge projector" is used. This projector has the nice property that the sum rule is satisfied, 
in the sense that the total electronic charges are conserved when summing up partial charges over all projected channels.

\par Specifically, one needs to define dual orbitals associated with the original atomic orbitals as 
\begin{equation}\label{eq:B2}
\tilde{\phi}_{\textbf{k}\mu}(\bfr) = \sum_{\nu} \phi_{\textbf{k} \nu}(\bfr) S_{\nu \mu }^{-1}\left( \textbf{k} \right) , 
\end{equation}
where
\begin{equation}
\phi_{\textbf{k} \nu}(\bfr) =  \sum_{\textbf{R}}  e^{i \textbf{k}\textbf{R}} \phi_{ \nu}^{\bfR} \left( \bfr \right) 
\end{equation}
is the Bloch summation of NAOs. And
$\phi_{ \nu}^{\bfR} \left( \bfr \right)= \phi_{\nu}\left( \textbf{r} - {\bm \tau}_{a} - \textbf{R} \right)$ denotes a NAO 
centering on the \textit{a}-th atom within the unit cell \textbf{R}. The orbital indices $\mu$ and $\nu$ are a combination of $\{a, l, \zeta, m\}$,  with $a$ labeling the atomic site, $l$, $m$ the angular and magnetic momentum,  and $\zeta$ the multiplicity (different radial functions) for a given $l$, respectively. Furthermore, $S_{\mu \nu}\left(\textbf{k}\right)$ is the overlap matrix in reciprocal space \cite{Li2016a}
\begin{equation}\label{eq:B3}
	S_{\mu \nu}\left(\textbf{k}\right) = 	\langle \phi_{ \textbf{k}\mu} |  \phi_{\textbf{k} \nu} \rangle
	= \sum_{\textbf{R}} e^{-i \textbf{k} \textbf{R}} \langle \phi_{\mu}^{\bfR} | \phi_{\nu}^{\textbf{0}} \rangle \, .
\end{equation}
Here we follow the convention that the extended Bloch orbitals are normalized within the BvK supercell cell and the real-space integration indicated by the braket 
goes over the supercell cell.
It can be readily shown that the dual and original Bloch orbitals satisfy the following biorthogonality relation
\begin{equation} \label{eq:B4}
	\langle \phi_{ \textbf{q}\mu} |  \tilde{\phi}_{\textbf{k} \nu} \rangle = \delta_{\textbf{k} \textbf{q}} \delta_{\mu \nu}\, .
\end{equation}

\par Making use of the dual orbitals, we define the projector used in the present work as 
\begin{equation}\label{eq:B5}
\begin{aligned}
\hat{P}_{I, m m^\prime}^{\sigma} =& \frac{1}{4N_{\textbf{k}}}\sum_{\textbf{k}} \left(  | \tilde{\phi}_{\textbf{k}, \beta  m^\prime} \sigma \rangle \langle \phi_{\textbf{k}, \beta  m} \sigma| \right. \\
& + | \phi_{\textbf{k}, \beta m^\prime} \sigma \rangle \langle \tilde{\phi}_{\textbf{k}, \beta  m} \sigma| 
  + | \tilde{\phi}_{\textbf{k}, \beta  m} \sigma \rangle \langle \phi_{\textbf{k}, \beta  m^\prime} \sigma| \\
& + \left. | \phi_{\textbf{k}, \beta m} \sigma \rangle \langle \tilde{\phi}_{\textbf{k}, \beta m^\prime} \sigma| \right)
\end{aligned}  ~,
\end{equation}
whereby the index $\beta$ groups together the indices $\{I, l, \zeta\}$ with $l$ and $\zeta$ belonging to the correlated channel of the correlated atom $I$. Note that this projector is 
slightly different from that introduced in the work of Han \textit{et al} \cite{Han2006}, which is essentially
an average of the first two terms in Eq.~(\ref{eq:B5}). The local occupation matrices yielded by the projector used in Ref.~\cite{Han2006} are
Hermitian but not necessarily real symmetric.
For convenience and numerical simplicity, we symmetrize the projector as is done in Eq.~(\ref{eq:B5}), and then
the resultant local occupation matrix is guaranteed to be real and symmetric. 
This is consistent with the feature that the ``on-site" global KS density matrix is also real and symmetric.
Inserting Eq.~(\ref{eq:B5}) into Eq.~(\ref{eq:projector2occupa}), we arrive at 
\begin{equation}\label{eq:B6}
\begin{aligned}
n_{I, m m^\prime}^{\sigma} =& \frac{1}{4N_{\textbf{k}}}  \sum_{\textbf{k}} \sum_{\mu} \left( S_{\beta m, \mu} \left( \textbf{k} \right) \rho_{\mu, \beta  m^\prime }^{\sigma} \left( \textbf{k}\right) \right. \\
+ &\left. \rho_{\beta m, \mu }^{\sigma} \left( \textbf{k}\right) S_{\mu, \beta m^\prime} \left( \textbf{k} \right) +  S_{\beta m^\prime , \mu} \left( \textbf{k} \right) \rho_{\mu, \beta  m}^{\sigma} \left( \textbf{k}\right) \right. \\
+& \left. \rho_{\beta m^\prime , \mu }^{\sigma} \left( \textbf{k}\right) S_{\mu, \beta m} \left( \textbf{k} \right) \right)
\, ,
\end{aligned} 
\end{equation}
where $\rho_{\mu \nu }^\sigma \left( \textbf{k} \right)$ is the spin-dependent KS density matrix
\begin{equation}\label{B7}
\rho_{\mu \nu }^\sigma \left( \textbf{k} \right) = \sum_{n} f_{n \textbf{k}}^\sigma c_{n \textbf{k}, \mu}^{\sigma} c_{n \textbf{k}, \nu}^{\sigma *} 
\end{equation}
with $c_{n \textbf{k}, \mu}^{\sigma}$ being the KS eigenvectors, satisfying the generalized orthogonality relationship,
\begin{equation}
    \sum_{\mu,\nu} c_{n \textbf{k}, \mu}^{\sigma} S_{\mu\nu}(\bfk) c_{n' \textbf{k}, \nu}^{\sigma\ast} = \delta_{nn'} \, .
    \label{eq:eigenvector_ortho}
\end{equation}

\par If we project the KS density matrix to all local atomic orbital channels and sum the traces of the resultant local occupation matrices up, 
i.e., requiring that $\beta, m$ go over all the basis indices $\{a, l, \zeta, m\}$, one then obtains
\begin{equation}\label{eq:B9}
\begin{aligned}
 &\sum_{\sigma, \nu} n_{\nu \nu}^\sigma \\
 &= \frac{1}{4N_{\textbf{k}}} \sum_{\sigma} \sum_{\textbf{k}} \sum_{\mu \nu}  \left( S_{\nu, \mu}\left( \textbf{k} \right) \rho_{\mu, \nu}^{\sigma } \left( \textbf{k}\right)  + \rho_{\nu, \mu }^{\sigma } \left( \textbf{k}\right) S_{\mu, \nu}\left( \textbf{k} \right) \right. \\
 & ~~~~\quad \left. + S_{\nu, \mu}\left( \textbf{k} \right) \rho_{\mu, \nu}^{\sigma } \left( \textbf{k}\right)  + \rho_{\nu, \mu }^{\sigma } \left( \textbf{k}\right) S_{\mu, \nu}\left( \textbf{k} \right) \right) \\
 & = \frac{1}{N_\bfk}\sum_{n,\sigma}\sum_{\bfk} f_{n\bfk}^\sigma \\ 
 &= N_{e}\, ,
\end{aligned} 
\end{equation}
where $N_e$ is the total number of electrons in one unit cell. In deriving Eq.~(\ref{eq:B9}), the orthogonality relationship Eq.~(\ref{eq:eigenvector_ortho}) is used.  Equation~(\ref{eq:B9}) is the above-mentioned sum rule satisfied by the ``Mulliken charge operator".

\subsection{Effective potential, force and stress}
\par To perform self-consistent DFT+$U$ calculations, and 
to enable structure relaxations, one needs to derive the expressions
of the effective single-particle potential and the forces and stresses corresponding to the DFT+$U$ energy functional. To this end, we first generalize the
single-site DFT+$U$ energy correction as given by Eq.~(\ref{eq:A10}) to the multi-site case,
\begin{equation}\label{eq:C1}
\Delta E_{\text{DFT+}U} = \frac{1}{2} \sum_{I} \bar{U}_{I}  \sum_{\sigma} \{ \text{Tr} \left( n_{I}^\sigma \right) -  \text{Tr} \left( n_{I}^\sigma n_{I}^\sigma \right)  \}\, ,
\end{equation}
where $\bar{U}_I=U_I-J_I$ is the effective interaction parameter on the correlated atom $I$. Again the isotropy of the interaction parameters is assumed.
The contribution of the energy correction $\Delta E_{\text{DFT+}U}$ to the KS effective potential operator is given by its derivative with respect to the $\bfk$-dependent density matrix operator,
\begin{equation}\label{eq:C2}
	 \hat{\rho}_{\textbf{k}}^\sigma  = \frac{1}{N_\bfk} \sum_{n} f^\sigma_{n \textbf{k}} |\psi_{n \textbf{k}}^\sigma \sigma \rangle \langle \psi_{n \textbf{k}}^{\sigma, \dagger} \sigma | ~.
\end{equation}
That is
\begin{equation}
\begin{aligned}\label{eq:C3}
\Delta\hat{V}_{\text{eff}}^\sigma \left( \textbf{k} \right) &= \frac{\delta \Delta E_{\text{DFT+}U}
}{\delta \hat{\rho}_{\textbf{k}}^\sigma} \nonumber \\
&=\sum_{I}\sum_{m m^\prime} \Delta V_{I, m m^\prime}^\sigma \hat{P}_{I, m m^\prime}^\sigma \left( \textbf{k} \right) 
\end{aligned} ~,
\end{equation}
where 
\begin{equation}\label{eq:C4}
\Delta V_{I, m m^\prime}^\sigma =  \bar{U}_{I} \left ( 1/2 \delta_{m m^\prime} -  n_{I, m m^\prime}^\sigma \right)  
\end{equation}
is the correction to the effective single-particle potential in the local subspace, and 
\begin{equation}\label{eq:C5}
\begin{aligned}
\hat{P}_{I, m m^\prime}^\sigma  \left( \textbf{k} \right) = &\frac{1}{4} \left( |\tilde{\phi}_{\textbf{k}, \beta  m^\prime} \sigma \rangle \langle \phi_{\textbf{k}, \beta  m} \sigma| +  | \phi_{\textbf{k}, \beta m^\prime}^\sigma \rangle \langle \tilde{\phi}_{\textbf{k}, \beta m} \sigma| \right.\\
& \left. + | \tilde{\phi}_{\textbf{k}, \beta  m} \sigma \rangle \langle \phi_{\textbf{k}, \beta  m^\prime} \sigma| +  | \phi_{\textbf{k}, \beta m} \sigma \rangle \langle \tilde{\phi}_{\textbf{k}, \beta m^\prime} \sigma| \right)\, .
\end{aligned}  
\end{equation}      
is the $\bfk$-dependent Mulliken projector. The matrix form of the effective potential within the full NAO basis set 
is given by
 \begin{equation}
  	\begin{aligned}
  		\Delta V_{\text{eff}, \mu \nu}^\sigma \left( \textbf{k} \right) 
  		=& \langle \phi_{\bfk\mu}| \Delta \hat{V}_{\text{eff}}^\sigma \left( \textbf{k} \right) |\phi_{\bfk\nu}  \rangle\\
  		=& \sum_{\textbf{R}} e^{-i \textbf{k} \textbf{R}} \langle \phi_\mu^\sigma \left( \textbf{R} \right) | \Delta \hat{V}_{\text{eff}}^\sigma \left( \textbf{k} \right) | \phi_{\nu}^\sigma \left( \textbf{0} \right) \rangle \\
  		=& \frac{1}{4} \sum_{I} \sum_{m m^\prime} \Delta V_{I, m m^\prime}^\sigma \left\{ S_{\beta m, \nu}\left( \textbf{k} \right) \delta_{\mu, \beta m^\prime} \right. \\
  		& + S_{\mu, \beta m^\prime}\left( \textbf{k} \right) \delta_{\beta m, \nu} + 
  		S_{\beta m^\prime, \nu}\left( \textbf{k} \right) \delta_{\mu, \beta m} \\
  		& + \left. S_{\mu, \beta m}\left( \textbf{k} \right) \delta_{\beta m^\prime, \nu} \right\} \, 
  	\end{aligned} 
\end{equation}
\label{eq:Delta_V_eff_matrix}
which is to be added to DFA Hamiltonian matrix to obtain the DFT+$U$ one.
As shown in Eq.~(\ref{eq:C4}), in the case of diagonal half-integer occupations the DFT+\textit{U} Hamiltonian reduces to standard DFAs. 

\par The contribution of the energy correction $\Delta E_{\text{DFT+}U}$ to the force on the $a$-th atom can be evaluated by its derivative with respect to the atomic coordinate ${\bm \tau}_a$
\begin{equation}\label{eq:C6}
\begin{aligned}
F_a =& \frac{d \Delta E_{\text{DFT+}U}}{d {\bm {\bm \tau}}_a}=\sum_{\sigma} \sum_{I} \sum_{m m^\prime}\Delta V_{I, m m^\prime}^\sigma \frac{d n_{I, m m^\prime}^{\sigma}}{d {\bm \tau}_a} ,
\end{aligned}
\end{equation}
which implies that the force due to the DFT+$U$ energy correction stems entirely from the change of the local occupation matrix in response to 
the atomic displacement. According to Eq.~(\ref{eq:B6}), the change of the local occupation matrix can arise either from the change of the overlap matrix $S$, or
from that of the KS density matrix $\rho$, namely, 
\begin{equation}\label{eq:deriv_n_split}
\begin{aligned}
   \frac{d n_{I, m m^\prime}^{\sigma}[S,\rho]}{d {\bm \tau}_a} = & \frac{\partial n_{I, m m^\prime}^{\sigma}[S,\rho]}{\partial S}\bigg|_\rho
   \frac{\partial S}{\partial {\bm \tau}_a} +  \\
   &  \frac{\partial n_{I, m m^\prime}^{\sigma}[S,\rho]}{\partial \rho}\bigg|_S 
   \frac{\partial \rho}{\partial {\bm \tau}_a}\, .
\end{aligned} 
\end{equation}
Thus, the force correction brought by DFT+$U$ is also split into two contributions. The first contribution, arising from the change of the overlap matrix, is given by
\begin{equation}
\begin{aligned}\label{eq:C7}
 &\sum_{\sigma} \sum_{I} \sum_{m m^\prime}\Delta V_{I, m m^\prime}^\sigma \frac{\partial n_{I, m m^\prime}^{\sigma}[S,\rho]}{\partial S}\bigg|_\rho \frac{\partial S}{\partial {\bm \tau}_a} \\
 & = \sum_{\sigma} \frac{1}{4N_{\textbf{k}}} \sum_{\textbf{k}}\sum_{I} \sum_{m m^\prime} \Delta V_{I, m m^\prime}^\sigma \sum_{\mu \in a} \\
 &\quad \left( \frac{\text{d} S_{\beta m, \mu} \left(\textbf{k} \right)}{\text{d} \bfr_{\beta m, \mu}} \rho_{\mu, \beta m^\prime}^{\sigma}\left( \textbf{k}\right)  +  \rho_{\mu, \beta m}^{\sigma *}\left( \textbf{k}\right) \frac{\text{d} S_{ \beta m^\prime, \mu}^* \left(\textbf{k} \right)}{\text{d} \bfr_{\beta m^\prime, \mu}} \right. \\
 &\quad +\left. \frac{\text{d} S_{\beta m^\prime, \mu}  \left(\textbf{k} \right)}{\text{d} \bfr_{\beta m^\prime, \mu}} \rho_{\mu, \beta m}^{\sigma}\left( \textbf{k}\right)  + \rho_{\mu, \beta m^\prime}^{\sigma *}\left( \textbf{k}\right)  \frac{\text{d} S_{\beta m, \mu}^* \left(\textbf{k} \right)}{\text{d} \bfr_{\beta m, \mu}} \right) \\
 & \quad  + \sum_{\sigma} \frac{1}{4N_{\textbf{k}}} \sum_{\textbf{k}}\sum_{I = a} \sum_{m m^\prime} \Delta V_{I, m m^\prime}^\sigma \sum_{\mu} \\
 & \quad \left( \frac{\text{d} S_{\mu, \beta m}^* \left(\textbf{k} \right)}{\text{d} \bfr_{\mu, \beta m}} \rho_{\beta m^\prime, \mu}^{\sigma *}\left( \textbf{k}\right)  + \rho_{\beta m, \mu}^{\sigma}\left( \textbf{k}\right)  \frac{\text{d} S_{ \mu, \beta m^\prime} \left(\textbf{k} \right)}{\text{d} \bfr_{\mu, \beta m^\prime}}  \right. \\
 &\quad +\left. \frac{\text{d} S_{\mu, \beta m^\prime}^* \left(\textbf{k} \right)}{\text{d} \bfr_{\mu, \beta m^\prime}} \rho_{\beta m,\mu}^{\sigma*}\left( \textbf{k}\right)  + \rho_{\beta m^\prime, \mu}^{\sigma}\left( \textbf{k}\right)  \frac{\text{d} S_{ \mu, \beta m} \left(\textbf{k} \right)}{\text{d} \bfr_{\mu, \beta m}} \right)
\end{aligned} ~,
\end{equation}
where $\bfr_{\nu, \mu} = {\bm \tau}_{a_\mu} - {\bm \tau}_{a_\nu}$ with $a_\nu$ and $a_\mu$ refer to the atoms that the NAO basis functions $\nu$ and $\mu$ are centering on, respectively. 
In the above derivation, we have used the Hermiticity of the overlap and density matrix, 
and the following relationship for the two-center integrals
\begin{equation}
 \frac{\partial \langle \phi_{\nu} \mid \phi_{\mu} \rangle}{\partial {\bm \tau}_a} = \left\{
 \begin{aligned}
  & \langle \phi_{\nu} \mid \frac{\phi_{\mu}}{d {\bm \tau_a}} \rangle = \frac{d \langle \phi_{\nu} \mid \phi_{\mu} \rangle}{d \bfr_{\nu, \mu}} , \mu \in a, \nu \notin a &\\
  & \langle \phi_{\mu} \mid \frac{d \phi_{\nu}}{d {\bm \tau_a}} \rangle =  \frac{d \langle \phi_{\mu} \mid \phi_{\nu} \rangle}{d \bfr_{\mu, \nu}}, \mu \notin a, \nu \in a  &\\
  & 0, ~~~~~~~~~~~~~~\mu \in a, \nu \in a ~\text{or}~\mu \notin a, \nu \notin a &
 \end{aligned} 
 \right. ~.
\end{equation}
The second part of the force consists in the contribution from the change of KS density matrix. Similarly it is denoted as $\frac{\partial n_{I, m m^\prime}^{\sigma}}{\partial {\bm \tau}_a}|_{S}$ which means that overlap matrix is fixed. Its contribution to the total force is given by
\begin{equation}\label{eq:C8}
 \begin{aligned}
 &\sum_{\sigma} \sum_{I} \sum_{m m^\prime}\Delta V_{I, m m^\prime}^\sigma \frac{\partial n_{I, m m^\prime}^{\sigma}[S,\rho]}{\partial \rho}\bigg|_S \frac{\partial \rho}{\partial {\bm \tau}_a} \\
 &= \sum_{\sigma} \frac{1}{4N_{\textbf{k}}} \sum_{\textbf{k}}\sum_{I} \sum_{m m^\prime} \Delta V_{I, m m^\prime}^\sigma \sum_{\mu} \\
 & \quad \left(  S_{\beta m, \mu} \left(\textbf{k} \right) \frac{ \text{d} \rho_{\mu, \beta m^\prime}^{\sigma}\left(\textbf{k} \right)}{\text{d} {\bm \tau}_a} + \frac{\text{d} \rho_{\beta m, \mu}^{\sigma}\left(\textbf{k} \right)}{\text{d} {\bm \tau}_a}  S_{ \mu, \beta m^\prime} \left(\textbf{k} \right) \right. \\
 & \quad +  \left.  S_{\beta m^\prime, \mu} \left(\textbf{k} \right) \frac{ \text{d} \rho_{\mu, \beta m}^{\sigma}\left(\textbf{k} \right)}{\text{d} {\bm \tau}_a} + \frac{\text{d} \rho_{\beta m^\prime, \mu}^{\sigma}\left(\textbf{k} \right)}{\text{d} {\bm \tau}_a}  S_{ \mu, \beta m} \left(\textbf{k} \right)  \right)  \\
 &= \frac{1}{N_{\textbf{k}}}\sum_{\textbf{k} \sigma} \sum_{\mu \nu} \Delta V_{\text{eff}, \nu \mu}^{\sigma} \left( \bfk\right) \frac{\text{d} \rho_{\mu \nu}}{\text{d} {\bm \tau}_{a}}\left( \bfk\right)\, .
 \end{aligned} 
\end{equation}
This term is the so-called orthogonality force: $\sum_{\mu \nu} H_{\nu \mu} \partial \rho_{\mu \nu}/ \partial {\bm \tau}_a$. Readers who are interested in the detailed derivation of this term are referred to Refs. \cite{Soler2002a, Li2016a}. However, this term in fact requires no additional treatment here since it has already been properly included in orthogonality force term via the usual DFA calculation procedure. 

\par The stress is defined as the derivative of the total energy with respect to the strain tensor. An efficient evaluation of the stress tensor is essential for
relaxing the shape and size of the unit cell of periodic systems. The stress contribution from the DFT+\textit{U} energy correction can be expressed as
\begin{equation}\label{eq:C9}
\begin{aligned}
\sigma_{\gamma \eta} =\frac{\partial \Delta E_{\text{DFT+}U}}{\partial \epsilon_{\gamma \eta}}=\sum_{\sigma} \sum_{I} \sum_{m m^\prime} V_{I, m m^\prime}^\sigma \frac{\partial n_{I, m m^\prime}^{\sigma}}{\partial \epsilon_{\gamma \eta}}
\end{aligned} ,
\end{equation}
where $\gamma$ and $\eta$ denote the Cartesian coordinate indices, and $\epsilon$ the strain tensor. As pointed out in Ref. \cite{Soler2002a}, stress calculations require
very little extra effort beyond multiplying the counterpart of the force by $\bfr_{\mu, \nu}^{\eta}$, where $\bfr_{\mu,\nu}$ is the vector connecting the atoms 
where the NAO basis 
functions $\phi_\mu$ and $\phi_\nu$ are centering on. Therefore, similar to the case of force calculations (cf. Eq.~\ref{eq:deriv_n_split}), the contributions tothe stress can also be decomposed into two parts. The first part $\frac{\partial n_{I, m m^\prime}^{\sigma}}{\partial \epsilon_{\gamma \eta}}|_\rho$, arising from the derivative of the overlap matrix with respect to
the strain tensor, is given by 
\begin{equation}\label{eq:C10}
\begin{aligned}
&\sum_{\sigma} \sum_{I} \sum_{m m^\prime} V_{I, m m^\prime}^\sigma \frac{\partial n_{I, m m^\prime}^{\sigma}}{\partial \epsilon_{\gamma \eta}}|_\rho \\
&=  \sum_{\sigma} \frac{1}{4N_{\textbf{k}}} \sum_{\textbf{k}}\sum_{I} \sum_{m m^\prime}  V_{I, m m^\prime}^\sigma \sum_{\mu} \\
& \left( \rho_{\mu, \beta m^\prime}^{\sigma}\left(\textbf{k} \right) \sum_{\textbf{R}} \frac{\text{d} S_{\beta m, \mu} \left(\textbf{R} \right)}{\text{d} \bfr_{\beta m, \mu}^{\gamma}} \bfr_{\beta m, \mu}^{\eta} e^{-i \textbf{k} \textbf{R}} \right.  \\
&  \quad +  \rho_{\beta m, \mu}^{\sigma}\left(\textbf{k} \right) \sum_{\textbf{R}} \frac{\text{d}  S_{ \mu, \beta m^\prime} \left(\textbf{R} \right)}{\text{d} \bfr_{\mu, \beta m^\prime}^{\gamma}}  \bfr_{\mu, \beta m^\prime}^{\eta} e^{-i \textbf{k} \textbf{R}}  \\
&  \quad + \rho_{\mu, \beta m}^{\sigma}\left(\textbf{k} \right) \sum_{\textbf{R}} \frac{\text{d} S_{\beta m^\prime, \mu} \left(\textbf{R} \right)}{\text{d} \bfr_{\beta m^\prime, \mu}^{\gamma}} \bfr_{\beta m^\prime, \mu}^{\eta} e^{-i \textbf{k} \textbf{R}} \\
& \quad + \left. \rho_{\beta m^\prime, \mu}^{\sigma}\left(\textbf{k} \right) \sum_{\textbf{R}} \frac{\text{d}  S_{ \mu, \beta m} \left(\textbf{R} \right)}{\text{d} \bfr_{\mu, \beta m}^{\gamma}}  \bfr_{\mu, \beta m}^{\eta} e^{-i \textbf{k} \textbf{R}} \right)\, .
 \end{aligned} 
\end{equation}
The second part of the stress is the counterpart of the orthogonality force term (cf.~Eq.~(\ref{eq:C8})). For the same reason as in force calculations, this
term has been included automatically in the total orthogonality stress in usual DFA calculations and requires no additional treatment. Hence, for brevity,
its explicit expression is not given here.

\subsection{Spin-orbit coupling}
\label{sec:formalism_soc}

\par In the discussion of the NAO-based DFT+$U$ formalism presented above, the SOC effect is neglected. Physically 
the SOC stems from the interaction between 
the intrinsic magnetic moment of the electrons and the magnetic field induced by their orbital angular momenta
and is a consequence of the relativistic effect. The magnitude of the SOC increases with the atomic number; for systems containing heavy
elements, such effect must be taken into account to obtain physically meaningful results. Below we discuss how SOC is incorporated in our
DFT$+U$ implementation.

\par When the SOC is present, the KS eigenstates $|\psi_{n \textbf{k}}\rangle$ become two-component spinors,
 \begin{equation}\label{eq:D1}
|\psi_{n \textbf{k}}\rangle=\sum_{\mu}\left(\begin{array}{c}
c_{n \mu, \textbf{k}}^{\uparrow} \\
c_{n \mu, \textbf{k}}^{\downarrow}
\end{array}\right)  | \phi_{\bfk\mu}\rangle
= |\psi_{n \textbf{k}}^{\uparrow} \rangle |\uparrow \rangle +  |\psi_{n \textbf{k}}^{\downarrow} \rangle  |\downarrow \rangle
\end{equation}
where |$\uparrow\rangle$ and |$\downarrow\rangle$ are the up- and down-channel spin states, and $\psi_{n \textbf{k}}^{\uparrow}(\bfr)$
and $\psi_{n \textbf{k}}^{\downarrow}(\bfr)$ are the associated spatial wave functions.
The two-component spinor can be regarded as a superposition state of its two components: state $|\Psi_{n \textbf{k}}^{\uparrow} \rangle$ multiplied with spin-up function and 
state $|\Psi_{n \textbf{k}}^{\downarrow} \rangle$ with spin-down function. The form of such two-component eigenstates is different from the one-component eigenstates $\Psi_{n \textbf{k}}^{\sigma}\left(\textbf{r}\right)$ of the non-SOC Hamiltonian, whereby the variables of spin $\sigma$ and position $\textbf{r}$ are independent. In the spirit of the method of separation of variables, the eigenstates $\Psi_{n \textbf{k}}^{\sigma}\left(\textbf{r}\right)$ is product of a function of the position $\Psi_{n \textbf{k}}\left(\textbf{r}\right)$ and a eigenstate of spin variables $\alpha_{n}\left( \sigma \right)$ with eigenvalue $\sigma$. Obviously the $|\Psi_{n \textbf{k}} \rangle$ in Eq.~(\ref{eq:D1}) cannot be reduced to such a form.

The DFT+$U$ energy functional is given by the expectation value of the second-quantized Hubbard Hamiltonian, i.e., Eq.~(\ref{eq:A2}), in the local subspace within the Hartree-Fock ground-state, given by
the Slater determinant that is formed by the lowest $N$ spinors. After some simple derivations, we obtain 
\begin{equation}\label{eq:D2}
	\begin{aligned}
		E_{U}=& \frac{1}{2} \sum_{I}\sum_{\{m\}} \sum_{\sigma \sigma^\prime}\left\langle m m^{\prime}\left|v_{s c}\right| m^{\prime \prime}  m^{\prime \prime \prime} \right\rangle n_{I,m m^{\prime \prime} }^{\sigma \sigma} n_{I, m^{\prime} m^{\prime \prime \prime}}^{\sigma^\prime \sigma^\prime}\\
		&-\frac{1}{2} \sum_{I} \sum_{\{m\}} \sum_{\sigma \sigma^\prime} \langle m m^{\prime} \left| v_{\mathrm{sc}}\right| m^{\prime \prime \prime} m^{\prime \prime}  \rangle n_{I, m m^{\prime \prime}}^{\sigma \sigma^\prime} n_{I, m^{\prime} m^{\prime \prime \prime}}^{ \sigma^\prime \sigma}
	\end{aligned} ~.
\end{equation}
The $n_{I,m m^{\prime} }^{\sigma \sigma^\prime}$ is the local occupation matrix within the SOC scheme that will be addressed later.  Following a similar procedure as Eq.~(\ref{eq:A5})$\sim$ Eq.~(\ref{eq:A7}), one can also introduce a
unitary transformation to Eq.~(\ref{eq:D2}) and then arrive at
\begin{equation}\label{eq:D3}
  \begin{aligned}
	E_{U}= & \frac{1}{2} \sum_{I} U_I \sum_{m m^\prime} \sum_{\sigma} \textbf{n}_{I, \sigma m} \textbf{n}_{I, -\sigma m^{\prime}} \\
	  & + \frac{1}{2} \sum_{I} \left( U_I-J_I \right) \sum_{m\neq m^\prime} \sum_{\sigma} \textbf{n}_{I, \sigma m} \textbf{n}_{I, \sigma m^{\prime}} 
  \end{aligned} ~.
\end{equation}
In this derivation, the $n_{I, m m^{\prime}}^{\sigma \sigma^\prime}$ is regarded as a local occupation matrix in the spin-orbit representation
rather a four-order tensor, i.e., $\sigma m$ is a compact index of spin-orbit. The $\textbf{n}_{I, \sigma m}$ is the diagonalized local occupation matrix
of $n_{I, m m^{\prime}}^{\sigma \sigma^\prime}$. 
For simplicity, here 
we have assumed that the ``off-diagonal blocks" of $n_{I, m m^{\prime}}^{\sigma \sigma^\prime}$ with $\sigma \ne \sigma'$
are much smaller in magnitude than the ``diagonal blocks" with $\sigma=\sigma'$, and that the eigenvalues obtained by diagonalizing 
$n_{I, m m^{\prime}}^{\sigma \sigma}$ separately for $\sigma=\uparrow,\downarrow$ don't differ appreciably from those obtained by diagonalizing the full local occupation
matrix. Test calculations indicate that this is a rather good approximation.
Subtracting the same double counting term as Eq.~(\ref{Eq:A8})
we get the energy correction within SOC scheme as 
\begin{equation}\label{eq:D4}
  \begin{aligned}
	\Delta E_{\text{DFT+}U} = & \frac{1}{2} \sum_{I} \left( U_I - J_I \right) \sum_{m \sigma} \left( \textbf{n}_{I, \sigma m} - \textbf{n}_{I, \sigma m} \textbf{n}_{I, \sigma m} \right) \\
	 =& \frac{1}{2} \sum_{I} \bar{U}_{I} \sum_{\sigma }\left[ \sum_{m} n_{I, m m}^{\sigma \sigma} \right. \\
	& \left. - \sum_{m m^\prime, \sigma^\prime} n_{I, m m^\prime}^{\sigma \sigma^\prime} n_{I, m^\prime m}^{\sigma^\prime \sigma} \right] .
  \end{aligned}
\end{equation}
The effective single-particle potential is given by 
\begin{equation}\label{eq:D5}
	V_{I, m m^\prime}^{\sigma \sigma^\prime} =  \bar{U}_{I} \left ( 1/2 \delta_{m m^\prime} \delta_{\sigma \sigma^\prime}-  n_{I, m m^\prime}^{\sigma \sigma^\prime} \right)  .
\end{equation}

\par Particularly, in this case, the Mulliken charge projector in Eq.~(\ref{eq:B5}) is generalized to a $2\times 2$ tensor in the spin space,
\begin{equation}\label{eq:D6}
\begin{aligned}
\hat{P}_{I, m m^\prime}^{\sigma, \sigma^\prime} =& \frac{1}{4 N_{\textbf{k}} }\sum_{\textbf{k}} \left(  | \tilde{\phi}_{\textbf{k}, \beta  m^\prime} \sigma^\prime \rangle \langle \phi_{\textbf{k}, \beta  m} \sigma| \right. \\
& + | \phi_{\textbf{k}, \beta m^\prime} \sigma^\prime \rangle \langle \tilde{\phi}_{\textbf{k}, \beta  m} \sigma | 
+ | \tilde{\phi}_{\textbf{k}, \beta  m} \sigma \rangle \langle \phi_{\textbf{k}, \beta  m^\prime} \sigma^\prime| \\
& \left. +  | \phi_{\textbf{k}, \beta m} \sigma \rangle \langle \tilde{\phi}_{\textbf{k}, \beta m^\prime} \sigma^\prime| \right)
\end{aligned}  ~.
\end{equation}
Thus the local occupation matrix in SOC case becomes
\begin{equation}\label{eq:D7}
  \begin{aligned}
	n_{I, m m^{\prime}}^{\sigma \sigma^\prime} =& \frac{1}{N_{\textbf{k}}}\sum_{n \textbf{k}} f_{n  \textbf{k}} \langle \psi_{n  \textbf{k}}| \hat{P}_{I, m m^\prime}^{\sigma, \sigma^\prime} | \psi_{n \textbf{k}} \rangle \\
	=& \frac{1}{4 N_{\textbf{k}}} \sum_{\textbf{k}} \sum_{\mu} \left( S_{\beta m, \mu} \left( \textbf{k} \right) \rho_{\mu, \beta  m^\prime }^{\sigma \sigma^\prime} \left( \textbf{k}\right) \right. \\
    &+ \left. \rho_{\beta m, \mu }^{\sigma \sigma^\prime} \left( \textbf{k}\right) S_{\mu, \beta m^\prime} \left( \textbf{k} \right) +  S_{\beta m^\prime , \mu} \left( \textbf{k} \right) \rho_{\mu, \beta  m}^{\sigma^\prime \sigma } \left( \textbf{k}\right) \right. \\
     &+ \left. \rho_{\beta m^\prime , \mu }^{\sigma^\prime \sigma } \left( \textbf{k}\right) S_{\mu, \beta m} \left( \textbf{k} \right) \right) 
  \end{aligned}  ~.
\end{equation}
Within the scheme of two-component spinors, 
the density matrix becomes a $2\times 2$ tensor in the spin space
\begin{equation}\label{eq:D8}
 \rho \left( \textbf{k} \right) =\left[\begin{array}{c c}
 \rho_{\mu \nu}^{\uparrow \uparrow} \left( \textbf{k} \right) & \rho_{\mu \nu}^{\uparrow \downarrow} \left( \textbf{k} \right) \\
 \rho_{\mu \nu}^{\downarrow \uparrow} \left( \textbf{k} \right) &\rho_{\mu \nu}^{\downarrow \downarrow}  \left( \textbf{k} \right)
 \end{array}\right] ,
\end{equation}
where
\begin{equation}\label{eq:D9}
\rho_{\mu \nu}^{\sigma \sigma^{\prime}} \left( \textbf{k} \right) = \sum_{n} f_{n \textbf{k}} c_{n \textbf{k}, \mu}^{\sigma} c_{n \textbf{k}, \nu}^{\sigma^{\prime} *} ~.
\end{equation}

\par Then the effective single-particle potential operator becomes
\begin{equation}\label{eq:D10}
	\hat{V}_{\text{eff}}^{\sigma \sigma^\prime} \left( \textbf{k} \right) 
	= \sum_{I} \sum_{m m^\prime} V_{I, m m^\prime}^{\sigma \sigma^\prime} \hat{P}_{I , m m^\prime}^{\sigma \sigma^\prime} \left( \textbf{k} \right) ,
\end{equation}
where
\begin{equation}\label{eq:D11}
\begin{aligned}
\hat{P}_{I, m m^\prime}^{\sigma, \sigma^\prime} \left( \textbf{k} \right)  =& 
\frac{1}{4}\left(  | \tilde{\phi}_{\textbf{k}, \beta  m^\prime} \sigma^\prime \rangle \langle \phi_{\textbf{k}, \beta  m} \sigma| 
+ |\phi_{\textbf{k}, \beta m^\prime} \sigma^\prime \rangle \langle \tilde{\phi}_{\textbf{k}, \beta  m} \sigma| \right. \\
&\left. + |\tilde{\phi}_{\textbf{k}, \beta  m} \sigma \rangle \langle \phi_{\textbf{k}, \beta  m^\prime} \sigma^\prime| 
  + |\phi_{\textbf{k}, \beta m} \sigma \rangle \langle \tilde{\phi}_{\textbf{k}, \beta m^\prime} \sigma^\prime| \right)
\end{aligned} ~.
\end{equation}

\subsection{$U$ and $J$ parameters determined from the Yukawa potential}
\label{sec:Yukawa}
\par As shown in the previous sections, the Hubbard \textit{U} and Hund \textit{J} are two key parameters in DFT+\textit{U} calculations. 
They can be taken as empirical parameters, or determined from pragmatic schemes like constrained DFT \cite{Dederichs1984,Norman1986,Gunnarsson1989,Anisimov1991}, constrained RPA \cite{Aryasetiawan2004,Miyake2008,Miyake2009,Sakuma2013}, or linear-response approach \cite{Cococcioni2005}. In interacting many-electron 
systems, the Coulomb interaction between electrons is screened, resulting weaker and often shorter-ranged effective interactions. 
In principle, one could directly model such a screened Coulomb potential, and directly use Eq.~(\ref{eq:A3}) to compute the $U$ and $J$ parameters. 
Previously, a simple form of such a screened potential -- the Yukawa potential has been employed in the literature, and demonstrated to work reasonably well
\cite{Bultmark2009, Norman1995, Wang2019}. In the present work, we follow such an approach and check how it works within our NAO-based DFT+$U$ scheme.


\par The Yukawa potential reads 
\begin{equation}\label{E4}
 v_\text{sc}\left( \mathbf{r}, \mathbf{r}^\prime \right) = \frac{e^{-\lambda | \mathbf{r}-\textbf{r}^\prime|}}{ | \mathbf{r}-\mathbf{r}^\prime|} 
 \end{equation}
 where $\lambda$ is a screening parameter. Compared to the bare Coulomb potential, the Yukawa potential decays exponentially fast to zero
 for large separations of two spatial points. Mathematically screened Coulomb interaction matrix elements based on the Yukawa potential given by Eq.~(\ref{eq:A3}) 
 can be decomposed into two parts, i.e., an angular integral part
involving spherical harmonics and a radial integral part called the Slater integrals~\cite{Liechtenstein1995}, namely,
\begin{equation}\label{E1}
 \langle m m^\prime \left| v_{sc} \right| m^{\prime \prime} m^{\prime \prime \prime} \rangle = \sum_{k=0}^{2l} a_k \left( m, m^\prime, m^{\prime \prime} m^{\prime \prime \prime} \right) F^{\left( k \right)} \, .
 \end{equation}
 In Eq.~(\ref{E1}), 
\begin{equation}\label{E2}
 \begin{aligned}
  &a_k \left( m, m^\prime, m^{\prime \prime} m^{\prime \prime \prime} \right) =\\
  & \frac{4 \pi}{2k+1} \sum_{q=-k}^{k} \int d \hat{\mathbf{r}} Y_{l m}^* \left( \hat{\mathbf{r}} \right) Y_{k q} \left( \hat{\mathbf{r}} \right) Y_{l m^{\prime \prime}} \left( \hat{\mathbf{r}} \right)   \\
  & \times \int d \hat{\mathbf{r}}^{\prime} Y_{l m^\prime}^* \left( \hat{\mathbf{r}}^\prime \right) Y_{k q}^* \left( \hat{\mathbf{r}}^\prime \right) Y_{l m^{\prime \prime \prime}} \left( \hat{\mathbf{r}}^\prime \right)
 \end{aligned}
 \end{equation}
 is the angular part which can be easily evaluated by Gaunt coefficients, and 
 \begin{equation}
 F^{\left( k \right)} = \int \int R^2_l \left( r \right) v_{sc}^{\left( k \right)} \left( r, r^\prime \right) R_l^2 \left( r^\prime \right) r^2 r^{\prime 2} dr dr^\prime 
 \label{eq:E3}
 \end{equation}
 is the radial part. The $v_{sc}^{\left( k \right)} \left( r, r^\prime \right)$ is the $k$-order coefficient of the expansion of $v_{sc}\left( \mathbf{r}, \mathbf{r}^\prime \right)$
 by spherical harmonics~\cite{Wang2019}, i.e.,
 \begin{equation}
  v_{sc}\left( \mathbf{r}, \mathbf{r}^\prime \right) = \sum_{k=0}^{\infty} \frac{4\pi}{2k+1} v_{sc}^{\left( k \right)} \left( r, r^\prime \right) 
  \sum_{q=-k}^{k} Y_{k q}^* \left( \hat{\mathbf{r}} \right) Y_{k q} \left( \hat{\mathbf{r}}^\prime \right) ~.
 \end{equation}
 In case of the Yukawa potential form, $F^{\left( k \right)}$ in Eq.~(\ref{eq:E3}) is further reduced to
 \begin{equation}\label{E5}
 \begin{aligned}
 &F^{\left( k \right)} = -\left( 2k+1 \right) \lambda \\
 &\times \int \int R^2_l \left( r \right) j_k \left(i \lambda r_<\right) h_k^{(1)}\left( i \lambda r_>\right) R_l^2 \left( r^\prime \right) r^2 r^{\prime 2} dr dr^\prime ,
 \end{aligned} 
\end{equation}
 where $j_k$ and $h_k^{(1)}$ are the spherical Bessel function and the spherical Hankel function of the first kind at order $k$, respectively. $r_>$ and $r_<$ are the smaller and the larger radius entering in the integral, i.e. $r_> = \text{max} \left( r, r^\prime\right)$ and $r_< = \text{min} \left( r, r^\prime\right)$. In this formulation, for a given set of local orbitals, 
 the matrix elements of the screened Coulomb potential as given by
 Eq.~(\ref{E1}), from which the $U$ and $J$ parameters can be extracted, depends only on 
 the screening parameter $\lambda$. 
 Thus, the problem of determining the $U$, $J$ values becomes one for determining $\lambda$.
 
\par In the theory of Thomas-Fermi model, the screening parameter $\lambda$ is a function of the charge density of the system,
\begin{equation}\label{E6}
	\lambda = 2 \left[ \frac{3\rho}{\pi}\right]^{1/6} ~,
\end{equation}
where $\rho$ is the electron density. 
In this work, we use the effective screening parameter $\bar{\lambda}$ which is obtained by averaging the space-dependent
screening parameter $\lambda(\bfr)$
\begin{equation}\label{E7}
	\bar{\lambda} = \frac{\int \lambda\left(\bfr\right) \rho\left(\bfr\right) d\bfr}{\int \rho\left(\bfr\right) d\bfr}  ~,
\end{equation}
where the $\rho\left(\bfr\right)$ is selected as the pseudo charge density in the KS self-consistent iteration.
This model provides us with a possible scheme of parameter-free DFT+\textit{U} with \textit{U} and \textit{J} determined in a self-consistent way. 

\par For practical DFT+\textit{U} calculations, determining the $U$ and $J$ values based on the Yukawa potential brings simplification. Firstly, in standard DFT+\textit{U} calculations, two parameters, i.e. Hubbard \textit{U} and Hund \textit{J}, need to be determined, while in the Yukawa-potential approach only one parameter $\lambda$ is needed. The screening parameter $\lambda$ can be evaluated self-consistently with the help of the Thomas-Fermi screening model. Thus a parameter-free DFT+\textit{U} scheme is in principle achievable. Secondly, for systems that have not been well investigated and no reference
results are available, it is highly nontrivial to obtain the appropriate \textit{U} and \textit{J} values. Under such
circumstances, a universal screening parameter $\lambda$ in this approach can provide initial information for further investigation. 
Thirdly, for accurate NAO-based calculations, there are often more than one radial function used for each $d/f$ angular momentum channel. 
This raises the question if the Hubbard $U$ correction need to be applied to all these $d/f$ function channels and how to determine the $U$,
$J$ parameters for each individual orbital. Based on Eq.~(\ref{E1}), the Yukawa potential approach allows one to conveniently determine 
the orbital-dependent $U$, $J$ parameters, from which one can readily decide the relevant orbital channels where the Hubbard $U$ correction
is necessary.

\section{Computational details}\label{sec:details}

\par Our DFT+$U$ implementation is carried out within the ABACUS code package \cite{Li2016a}. In the present work,  we use the SG15 optimized norm-conserving 
Vanderbilt (ONCV) multi-projector pseudo-potentials \cite{Hamann2013, Schlipf2015, Scherpelz2016} to describe the ion cores, and optimized double-$\zeta$ plus polarization
(DZP) atomic basis sets \cite{Chen2010,Lin2021} to expand the Kohn-Sham eigenfunctions. Within DZP, $4s2p2d1f$ atomic functions with a cutoff radius of 9 Bohr are used
for the TM atoms and $2s2p1d$ atomic functions with $r_\text{cutoff}$ of \SI{7}{Bohr} for oxygen (O) atoms. 
In certain occasions, the triple-$\zeta$ plus polarization (TZDP) basis set with $5s3p3d2f$ for TMs
and $3s3p2d$ for O are also used.
For Brillouin zone sampling,  a $\Gamma$-inclusive $6\times6\times6$ \textbf{k}-mesh is used. 
In band structure calculations, we set up \textbf{k}-point paths explicitly along specified high-symmetry directions of the Brillouin zone. We consider ten high-symmetry 
points in \textbf{k}-space for the rhombohedral Bravais lattice of  type-\uppercase\expandafter{\romannumeral2} antiferromagnetic (AFM) structures (e.g. late TM monoxides 
MnO, FeO, CoO and NiO). The corresponding \textbf{k}-path is $\Gamma-L-B_1|B-Z-\Gamma-X|Q-F-P_1-Z|L-P$ \cite{Setyawan2010}. 
As for the case of simple tetragonal IrO$_2$, the $\bfk$-path is $\Gamma-X-M-\Gamma-Z-R-A-M$ \cite{Panda2014}. For all calculations, the Perdew-Burke-Ernzerhof (PBE) generalized gradient approximation (GGA) 
is used as the exchange-correlation functional \cite{Perdew1996} at the level of DFA, i.e., the PBE+$U$ scheme is employed in our test calculations below. 

\par To validate our scheme and implementation we make comparisons to several widely used softwares, including the LAPW method based code WIEN2k, Vienna \textit{ab initio} simulation package (VASP) which is based on the PAW method, the Quantum-ESPRESSO (QE) package which uses pseudopotential and plane-wave basis set \cite{Giannozzi2009, Cococcioni2005}, and LCAO based OpenMX code \cite{Ozaki2003, Ozaki2004, Ozaki2004a, Han2006}. For all calculations, we use the isotropic version of the simplified DFT+\textit{U} method, in which only the difference between Hubbard \textit{U} and Hund \textit{J} matters \cite{Dudarev1998a}. For all WIEN2k based calculations, we set the muffin-tin radii of TM atoms as \SI{2.1}{Bohr} and that for O atoms as \SI{1.4}{Bohr}. The convergence of the calculations is controlled by $R_{\text{MT}} \times K_{\text{max}} = 7.0$. In VASP calculations, we set the energy cutoff as \SI{700}{\electronvolt}, and non-spherical contribution in gradient corrections inside the PAW sphere is included (LASPH=.True.). In order to make comparisons, we use the same SG15 ONCV multi-projector pseudopotentials in QE calculations. The kinetic energy cutoffs for wavefunctions and charge density  are 60 and \SI{480}{Ry} respectively. For OpenMX results, we simply cite those reported in Ref.~\cite{Han2006}. 

\par The introduction of multiple radial functions (multiple-$\zeta$) with the same angular
momentum to the basis set in the LCAO framework makes it intricate to define the projector, 
and hence the local subspace. The definition of the projector in section \ref{sec:mulliken}
indicates that the correlated sub-shell is not uniquely specified with only the correlated site $I$ and angular
momentum $l$, because there is still a $\zeta$ index. In the present work, the DZP basis sets
we are using contain two $d$-type functions and there is naturally the question regarding how to define the
$\beta$ index when constructing the projector (cf. Eq.~(\ref{eq:B5}) and (\ref{eq:D2})). In our current
procedure \cite{Chen2010,Lin2021} to optimize the NAO basis sets, the first generated $d$ function of TMs is
most localized and represents best the local correlated subspace around the Fermi level. Also, the estimated $U$ and
$J$ parameters for these orbitals based on the Yukawa potential as described in Sec.~\ref{sec:Yukawa} also agree fairly well with those reported in the literature. 
In contrast, the second $d$ function is rather delocalized and contributes little to the correlated subspace. 
As such, in the calculations reported below, the first $d$ orbital is used to construct the projector.
Further discussions about our choice and its efficacy will be discussed in section \ref{sec:4e} and appendix \ref{sec:appendix}.

\section{Results and discussion}\label{sec:results}
\subsection{Band gaps and magnetic moment}\label{sec:4a}
\par One of the most noteworthy successes of the DFT+\textit{U} approach is the reproducing of the observed insulating behavior of TMOs. In the standard DFT scheme, LDA and GGA either predict conducting behavior or yield gaps that are much smaller than the experimental values. In TMOs, correlated valence \textit{d} electrons tend to be localized due to strong on-site Coulomb interaction. Local and semi-local DFAs fail to describe this localization behavior. This deficiency is largely rectified by DFT+$U$. 
Here, we apply our NAO-based DFT+$U$ implementation to prototypical Mott insulators MnO, FeO, CoO and NiO, and check how the obtained band gaps vary with the $\bar{U}$ value. 
We also compare our results with those produced by established computer codes, whereby the validity of our implementation can be demonstrated. 

\par In our calculations, the NaCl-type structure is assumed for the four late TMOs -- MnO, FeO, CoO and NiO -- where the chosen lattice constants are
 \SI{4.445}{\angstrom}, \SI{4.334}{\angstrom}, \SI{4.254}{\angstrom} and \SI{4.171}{\angstrom}, respectively \cite{Tran2006}. For all these four systems 
we consider the type-\uppercase\expandafter{\romannumeral2} AFM structure, where the AFM order is along the $\left(111\right)$ direction
\cite{Cococcioni2005, Wang2016, Wang2019}. The SOC effect is not considered for these systems.

\par Table \ref{table1} presents the calculated gaps as a function of the effective on-site Coulomb interaction parameter $\bar{U}$. Without the $U$ correction, PBE yields zero gaps for FeO and CoO, and significantly underestimates the band gaps for MnO and NiO. 
When the $U$ correction is applied, the obtained  band gap opens up and steadily increases with an increasing $\bar{U}$ value, 
as expected. 
When $\bar{U}$ reaches a physically meaningful value
of about 6 eV, our calculated PBE+$U$ band gaps show a fairly good agreement with the experimental values for all four TMOs.
\begin{table}[ht]
	\centering
	\caption{ Band gaps (in eV) of MnO, FeO, CoO and NiO as a function of effective on-site Coulomb energy $\bar{U}$ (in eV). The experimental values
	are presented in the last row.}
	\begin{ruledtabular}
	\begin{tabular}{p{1.5cm}<{\centering} p{1.5cm}<{\centering} p{1.5cm}<{\centering} p{1.5cm}<{\centering} p{1.5cm}<{\centering}}
		$\bar{U}$ (eV)  & MnO  & FeO &CoO & NiO \\
		\colrule
		0.0  & 1.13      & 0.00     & 0.00     & 0.82   \\
		
		2.0  & 1.65      & 0.76     & 1.84   & 1.88    \\
		
		4.0  & 1.98      & 2.44     & 2.57     & 2.60   \\
		
		6.0  & 2.18      & 2.62     & 3.07     & 3.21   \\
		
		Expt. & 3.6-3.8$^{\text{a}}$  &  2.4$^{\text{b}}$    &2.4$^{\text{c}}$   &   4.0$^{\text{d}}$, 4.3$^{\text{e}}$         \\
	\end{tabular}
    \label{table1}
	\end{ruledtabular}
\begin{tablenotes}
	\item[1] $^\text{a}$Reference \cite{Messick1972} 
	\item[2] $^\text{b}$Reference \cite{Bowen1975}
	\item[3] $^\text{c}$Reference \cite{Powell1970}
	\item[4] $^\text{d}$Reference \cite{Sawatzky1984}
	\item[5] $^\text{e}$Reference \cite{Hufner1984}
\end{tablenotes}
\end{table}

\par
\begin{figure*}[htb]
	\graphicspath{{figures/}}
	\centering
	\subfigure{
		\centering
		\begin{minipage}[htb]{0.45\linewidth}
			\centering
			\includegraphics[width=\linewidth]{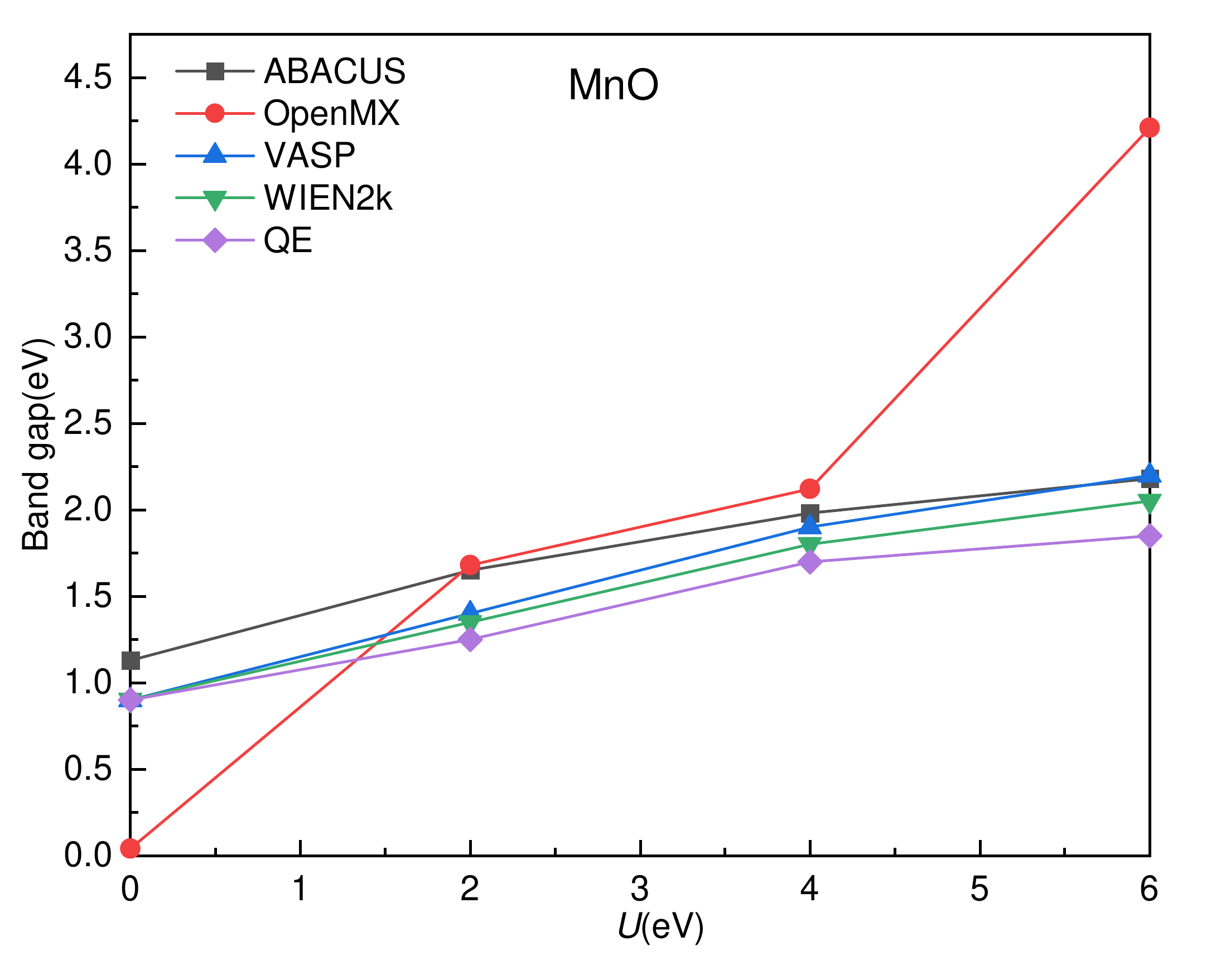}
			\includegraphics[width=\linewidth]{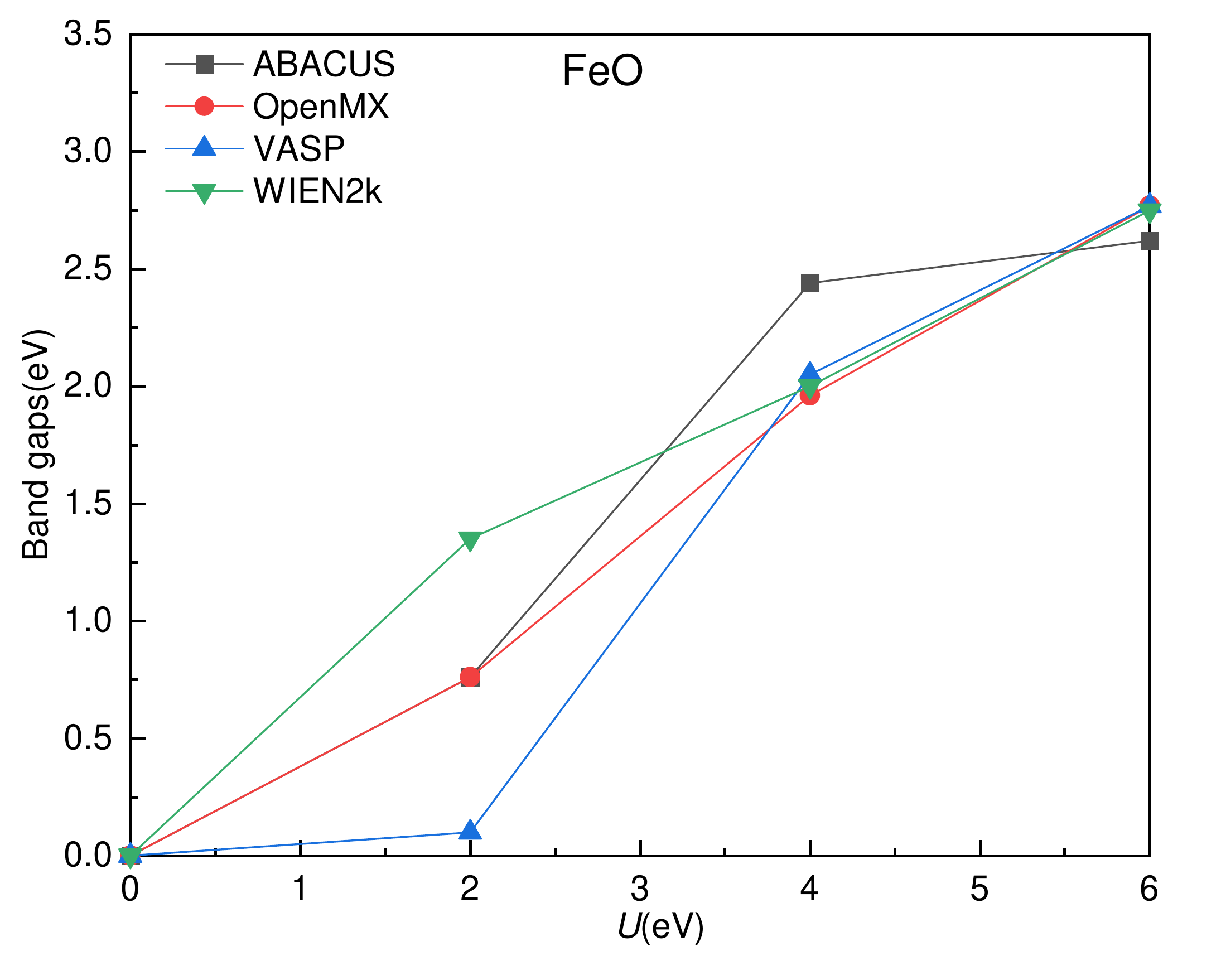}
		\end{minipage}
	}
	\hspace{5mm}
	\subfigure{
		\centering
		\begin{minipage}[htb]{0.45\linewidth}
			\centering
			\includegraphics[width=\linewidth]{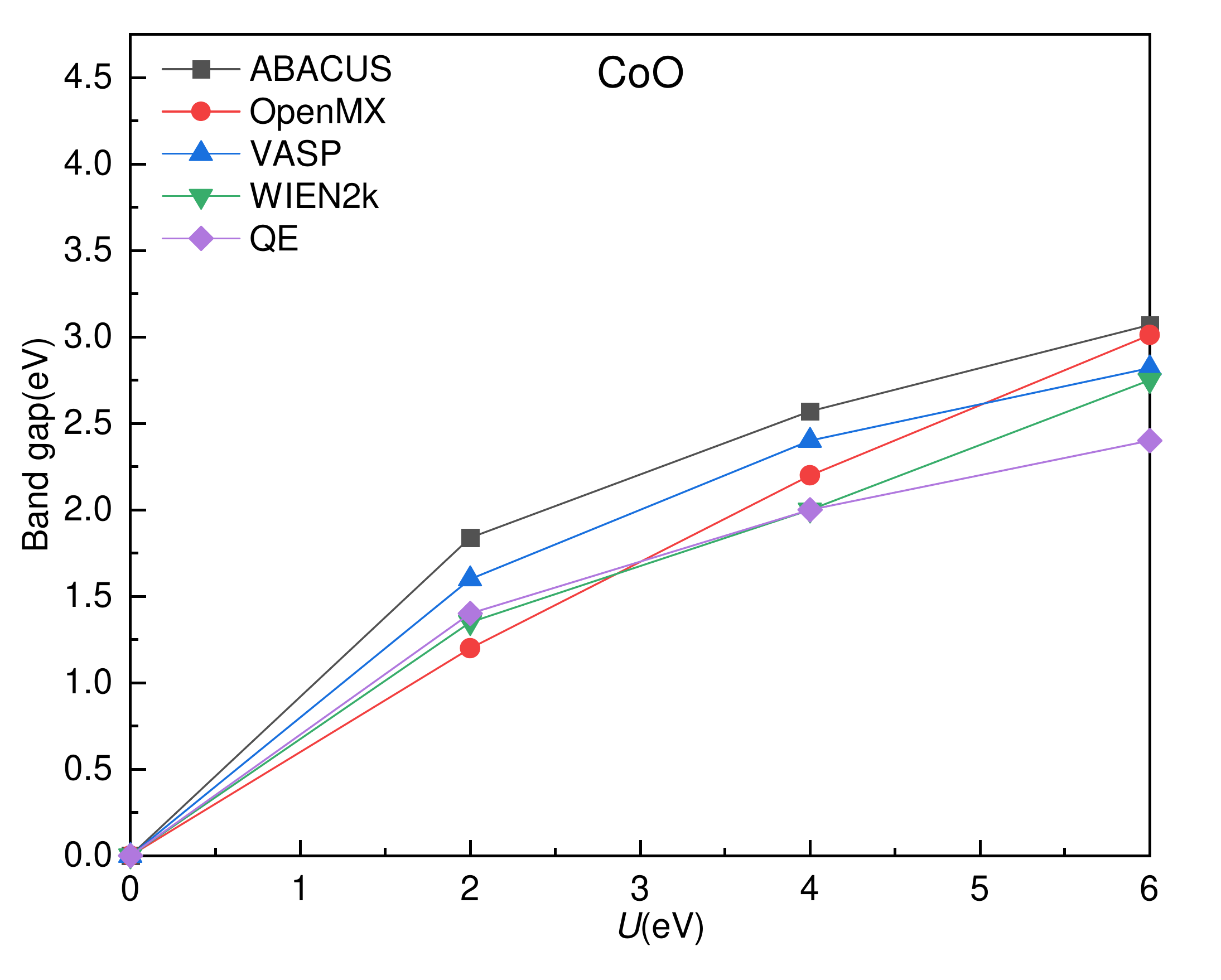}
			\includegraphics[width=\linewidth]{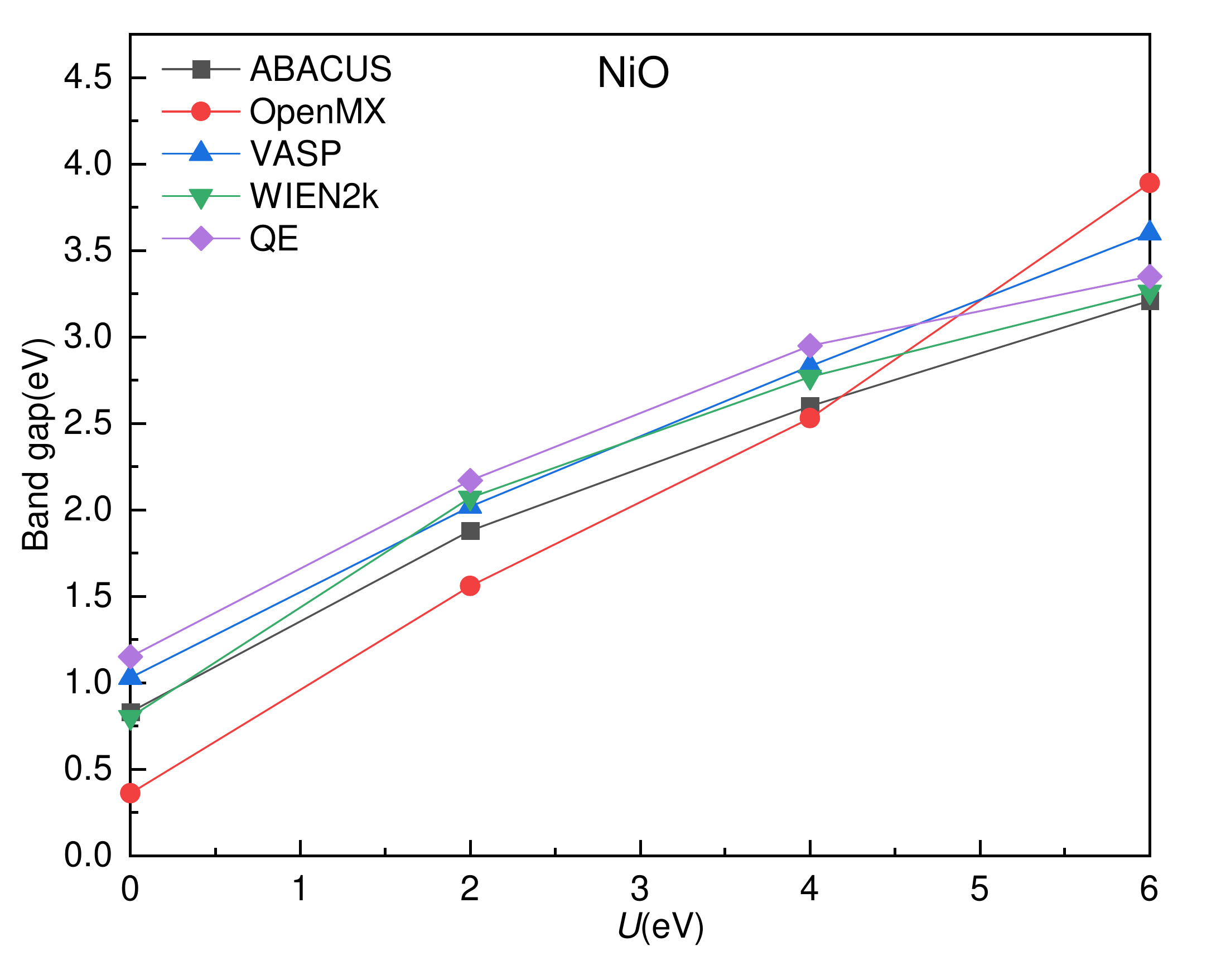}
		\end{minipage}
	}
	\caption{Band gaps of MnO, FeO, CoO and NiO as a function of $\bar{U}$ as calculated by different computer codes. The rhombohedral unit cell with  type-\uppercase\expandafter{\romannumeral2} AFM structure is used. For FeO, we couldn't obtain stable self-consistent results for all $\bar{U}$ values with QE,
	and hence the QE results are not included in FeO panel. The LDA+$U$ results taken from Ref.~(\cite{Han2006}) are presented for OpenMX, whereas for all other codes the PBE+$U$ results are presented.} 
	\label{Fig.1}
\end{figure*}

 Since different projectors and numerical frameworks are used in different computer codes, the calculated DFT+$U$ band gaps could differ noticeably even with the same
$U$ value. However, the results should be qualitatively similar. For instance, the calculated DFT+$U$ band gaps should follow a similar trend as the $U$ value increases.
To check this, in Fig.~\ref{Fig.1} we present the band gaps for the four TMOs as a function of $\bar{U}$, as obtained by five computer codes: ABACUS, OpenMX, VASP, WIEN2k,
and QE. 
The OpenMX results for all the four systems are taken from Ref. \cite{Han2006}, where LDA functional was used. For all other codes, the PBE functional is used in
DFT+$U$ calculations. We don't expect using PBE instead of LDA will produce significant difference for the curves presented in Fig.~\ref{Fig.1}.
For MnO, the results of VASP, WIEN2k and QE are taken from Ref. ~\cite{Wang2016}, whereas the results for other TMOs are calculated in the present work. 
Figure~\ref{Fig.1} indicates the band gaps obtained from different codes follow the same trend as $\bar{U}$ increases, but the obtained band gaps can differ by 
as large as 1 eV for the same  $\bar{U}$ value. This means that
the ``best" $\bar{U}$ value to reproduce the experimental results varies for different numerical frameworks. Compared to other codes, ABACUS displays a rather ``normal" behavior,
which is a strong evidence of the validity of our projector scheme and numerical implementations.


\par The local magnetic moment of TM atoms within MnO, FeO, CoO and NiO are also calculated by ABACUS and the results are presented in Table \ref{table2}. 
Table \ref{table2} shows the expected trend that the local magnetic moments get enhanced with increasing $\bar{U}$. 
With physical $\bar{U}$ values, the obtained PBE+$U$ magnetic moments show an overall good agreement with  experimental results and previously reported theoretical results \cite{Anisimov1991, Han2006}. 
\begin{table}[ht]
	\centering
	\caption{ Magnetic moments (in $\mu_B$) of TM atoms within the four TMOs as a function of effective on-site Coulomb energy $\bar{U}$. }
	\begin{ruledtabular}
	\begin{tabular}{p{1.5cm}<{\centering} p{1.5cm}<{\centering} p{1.5cm}<{\centering} p{1.5cm}<{\centering} p{1.5cm}<{\centering}}
		$\bar{U}$ (eV)  & MnO  & FeO & CoO & NiO \\
		\colrule
		0.0  & 4.39      & 3.44   & 2.47    & 1.21   \\
		
		2.0  & 4.64      & 3.60   & 2.56     & 1.53    \\
		
		4.0  & 4.74      & 3.70   & 2.69    & 1.64   \\
		
		6.0  & 4.80      & 3.77   & 2.74    & 1.71   \\
		
		Expt. & 4.58$^{\text{a}}$, 4.79$^{\text{b}}$  & 3.32$^{\text{c}}$  & 3.8$^{\text{c}}$, 3.35$^{\text{d}}$   &  1.90$^{\text{a}}$, 1.77$^{\text{b}}$   \\
        Theo. & 4.61$^{\text{e}}$  & 3.62$^{\text{e}}$  & 2.63$^{\text{e}}$   &  1.69$^{\text{e}}$,  1.74$^{\text{f}}$  \\
	\end{tabular}
   \label{table2}
	\end{ruledtabular}
	\begin{tablenotes}
		\item[1] $^\text{a}$Reference \cite{Cheetham1983}
		\item[2] $^\text{b}$Reference \cite{Fender1968}
		\item[3] $^\text{c}$Reference \cite{Roth1958}
		\item[4] $^\text{d}$Reference \cite{Khan1970} 
		\item[5] $^\text{e}$Reference \cite{Anisimov1991}
		\item[6] $^\text{f}$Reference \cite{Han2006} ($U$=\SI{6.0}{\electronvolt})
	\end{tablenotes}
\end{table}

\subsection{Projected density of states and band structures} \label{sec:4b}
\par
\begin{figure*}[htb]
	\graphicspath{{figures/}}
	\centering
		\subfigure{
			\centering
			\begin{minipage}[htb]{0.4\linewidth}
				\centering
				\includegraphics[width=\linewidth]{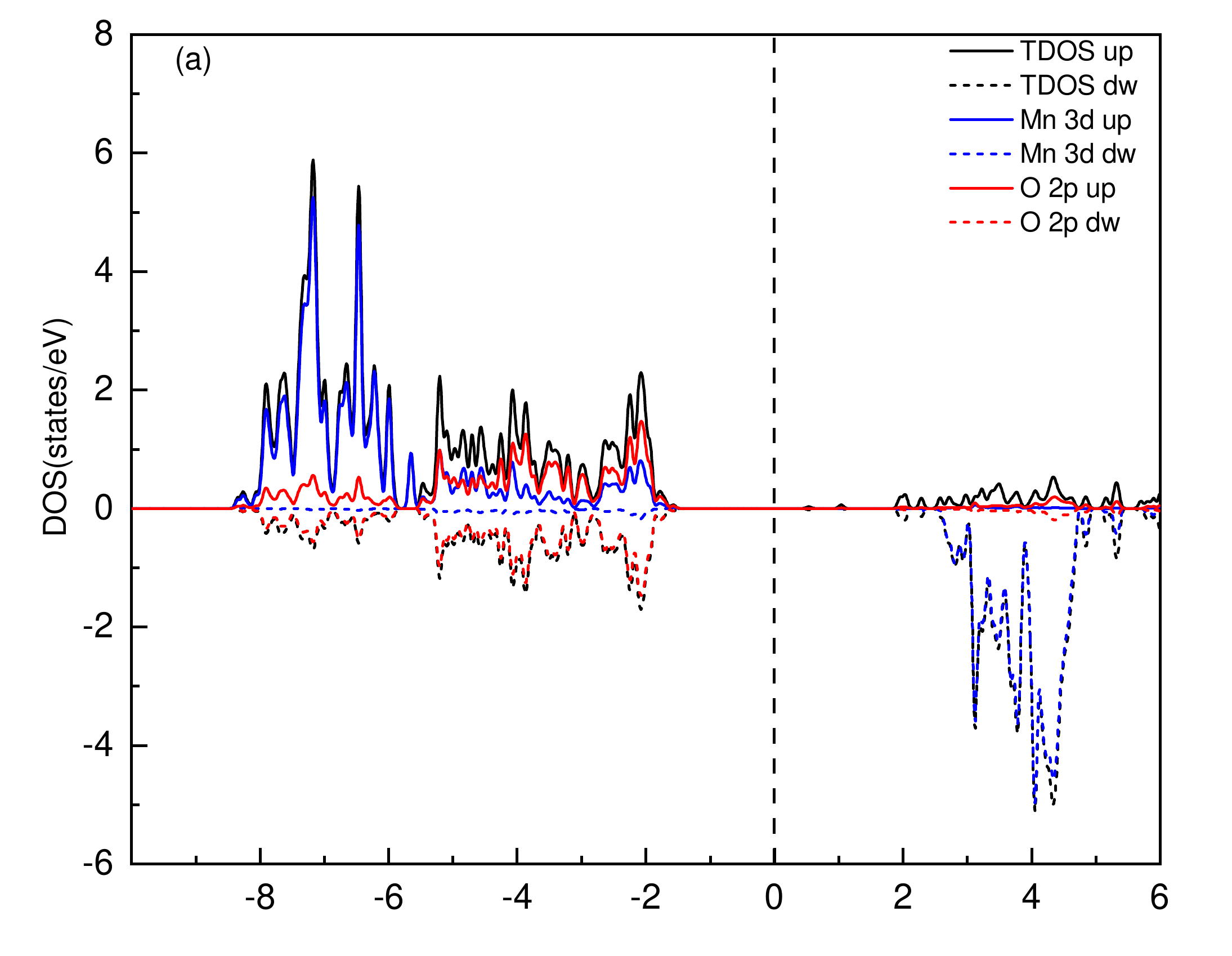}
				\includegraphics[width=\linewidth]{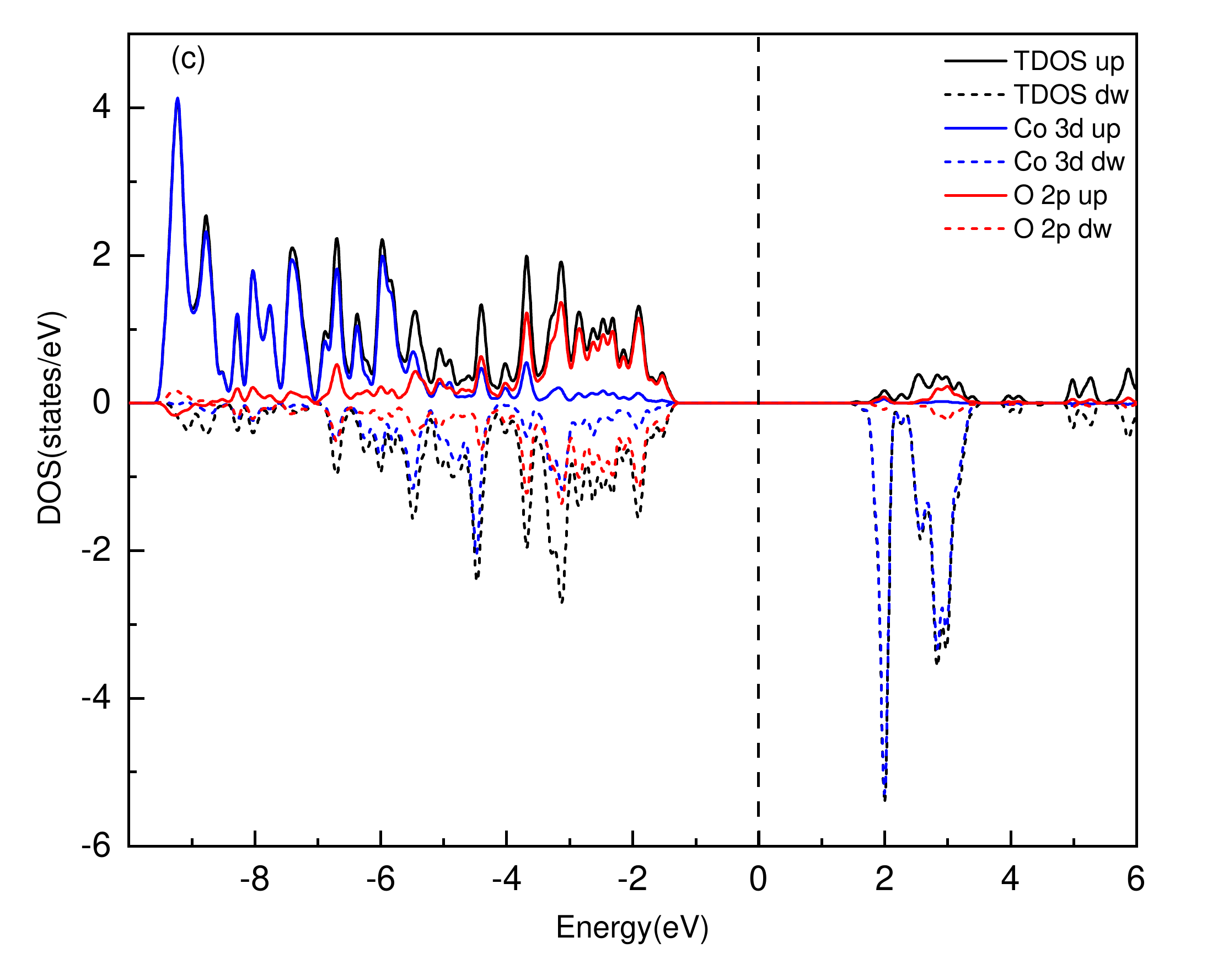}
			\end{minipage}
		}
		\subfigure{
			\centering
			\begin{minipage}[htb]{0.4\linewidth}
				\centering
				\includegraphics[width=\linewidth]{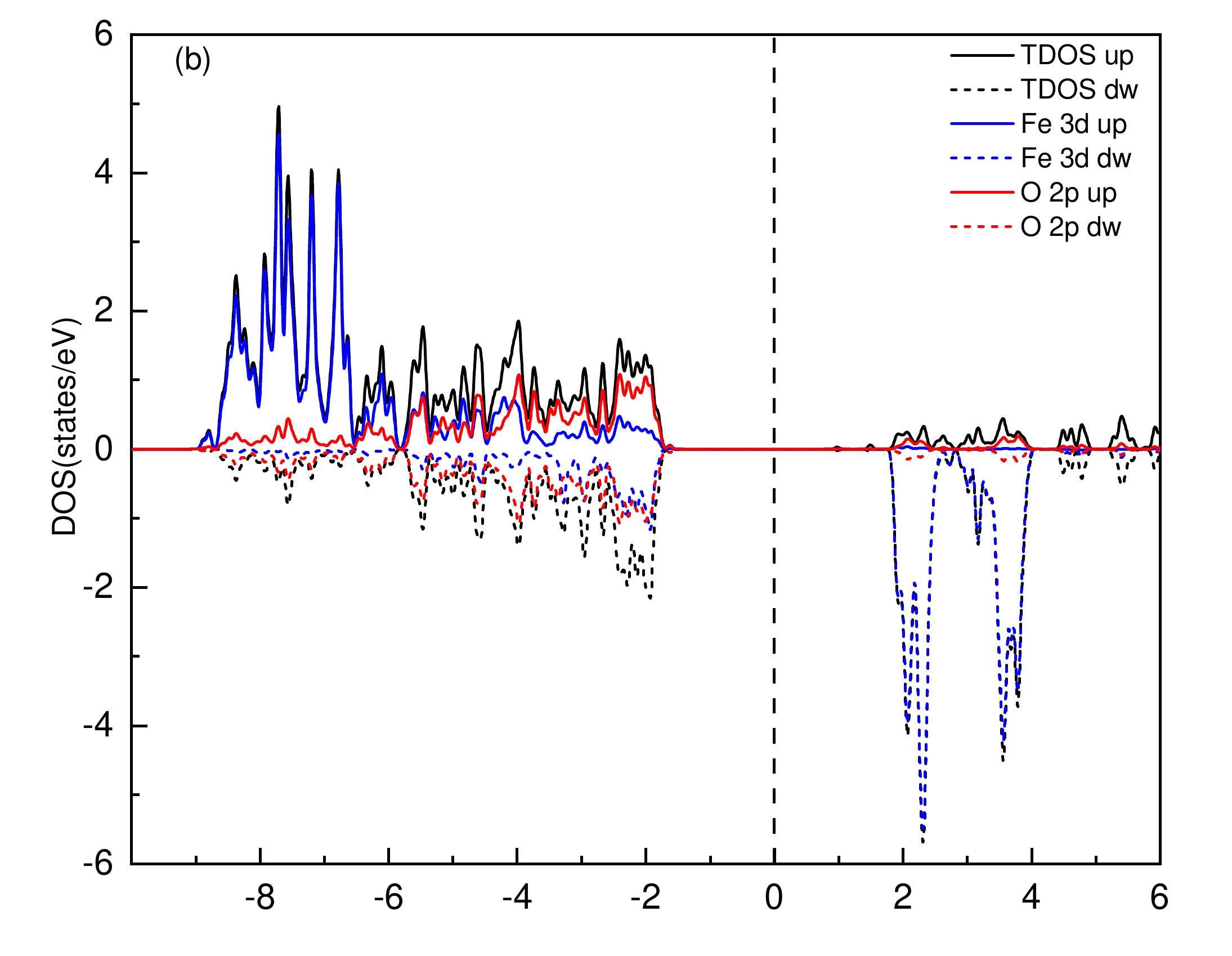}
				\includegraphics[width=\linewidth]{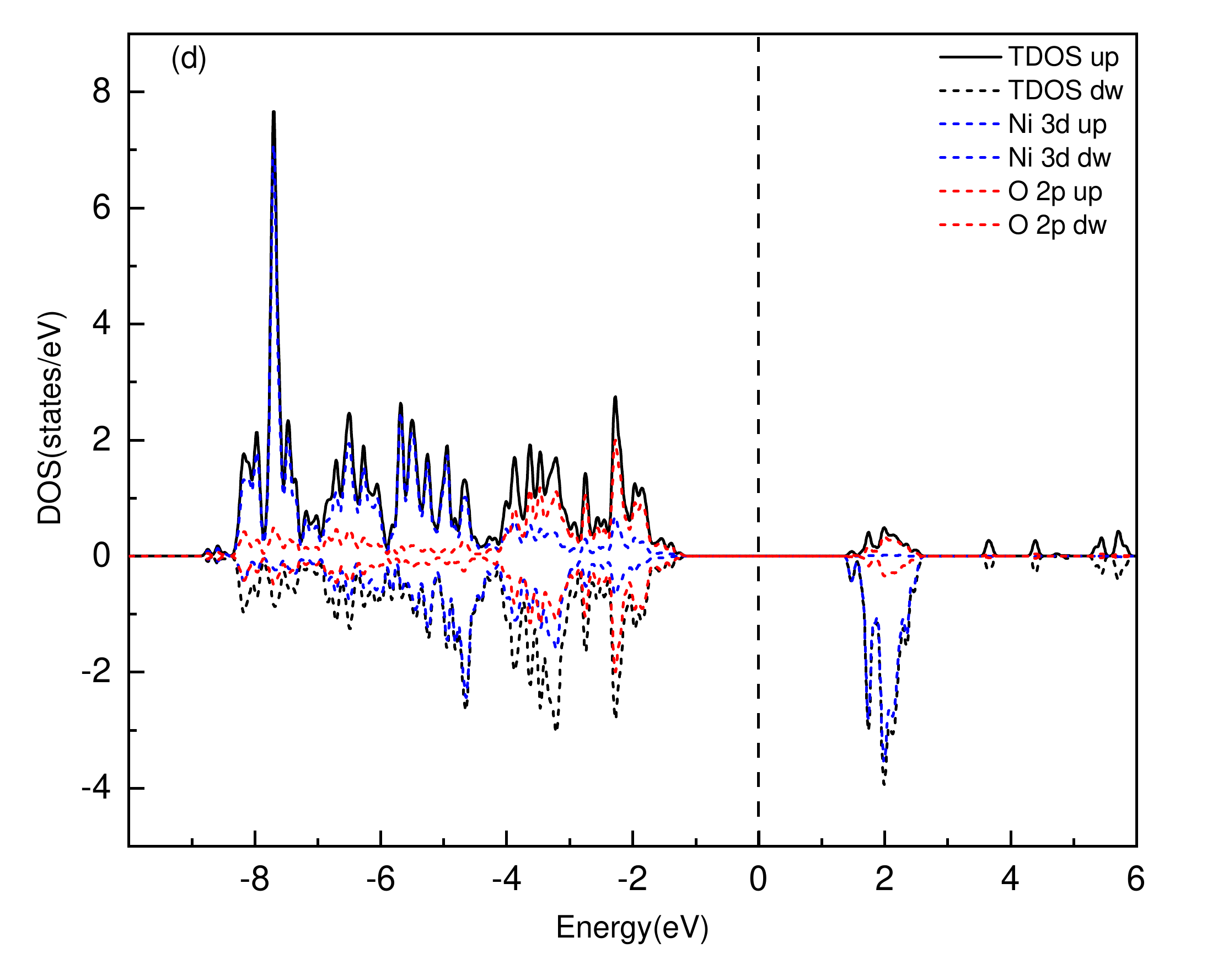}
			\end{minipage}
		}
	\caption{The TDOS and PDOS of spin-up polarized TM atom and its nearest O atom of MnO, FeO, CoO and NiO calculated with $\bar{U}$ of \SI{5}{\electronvolt} are depicted in (a), (b), (c), and (d), respectively. The zeroes of the energy axes are set to be the Fermi level (vertical dashed line). The upper and lower panels correspond to the 
	up-spin and down-spin channels.}
	\label{Fig.2}
\end{figure*}

\begin{figure*}[htb]
	\graphicspath{{figures/}}
	\centering
	\subfigure{
		\centering
		\begin{minipage}[htb]{0.4\linewidth}
			\centering
			\includegraphics[width=\linewidth]{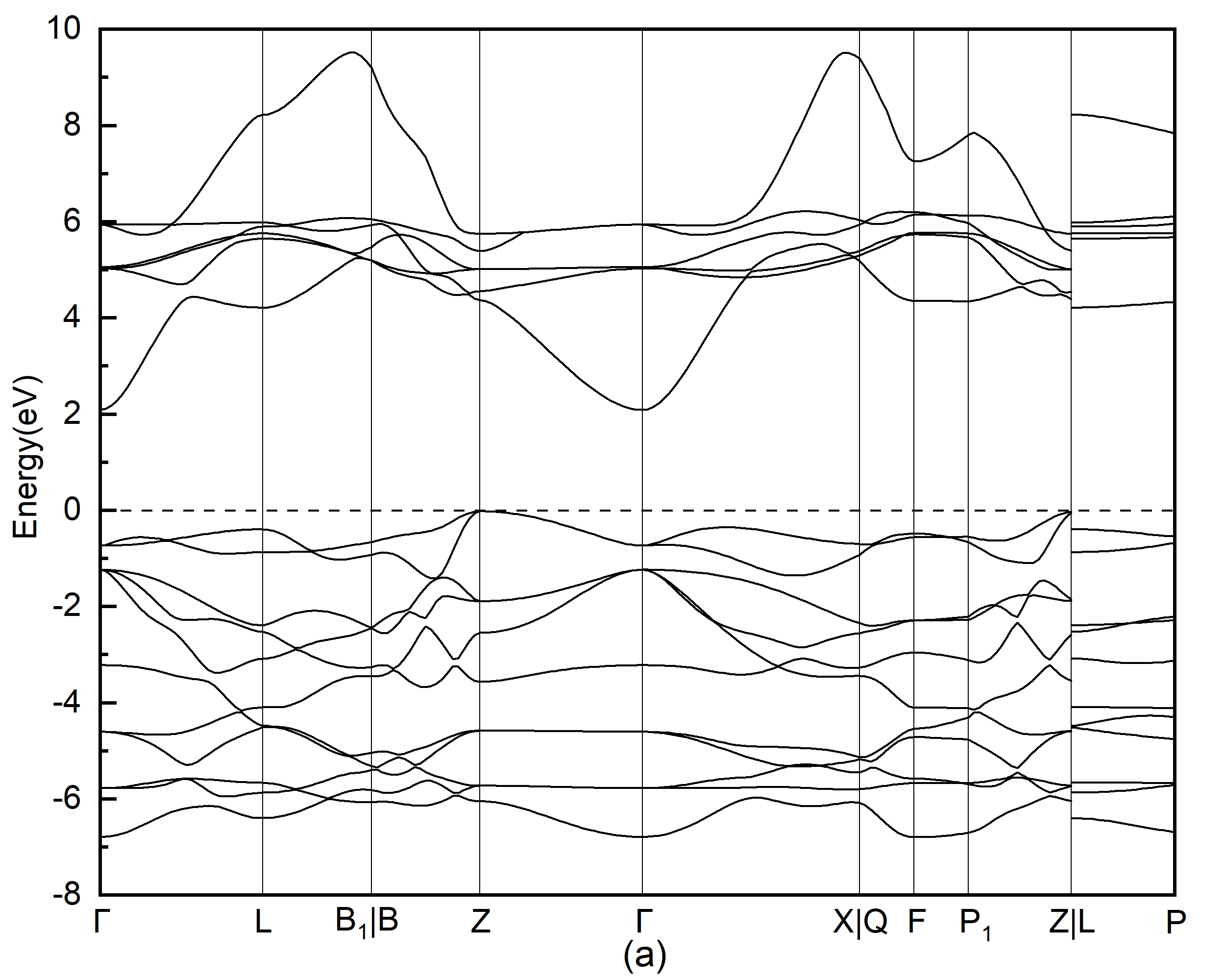}
			\includegraphics[width=\linewidth]{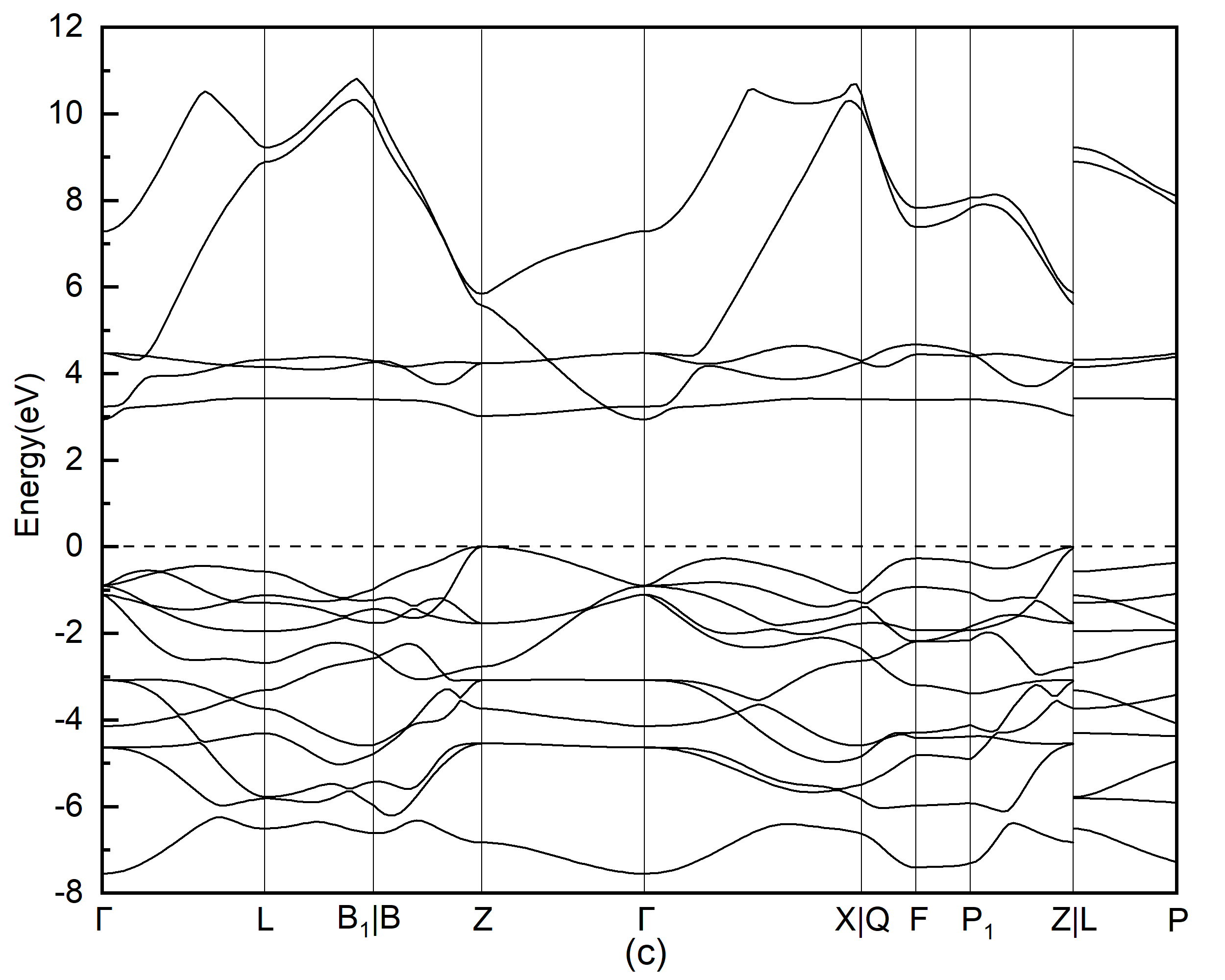}
		\end{minipage}
	}
	\hspace{5mm}
	\subfigure{
		\centering
		\begin{minipage}[htb]{0.4\linewidth}
			\centering
			\includegraphics[width=\linewidth]{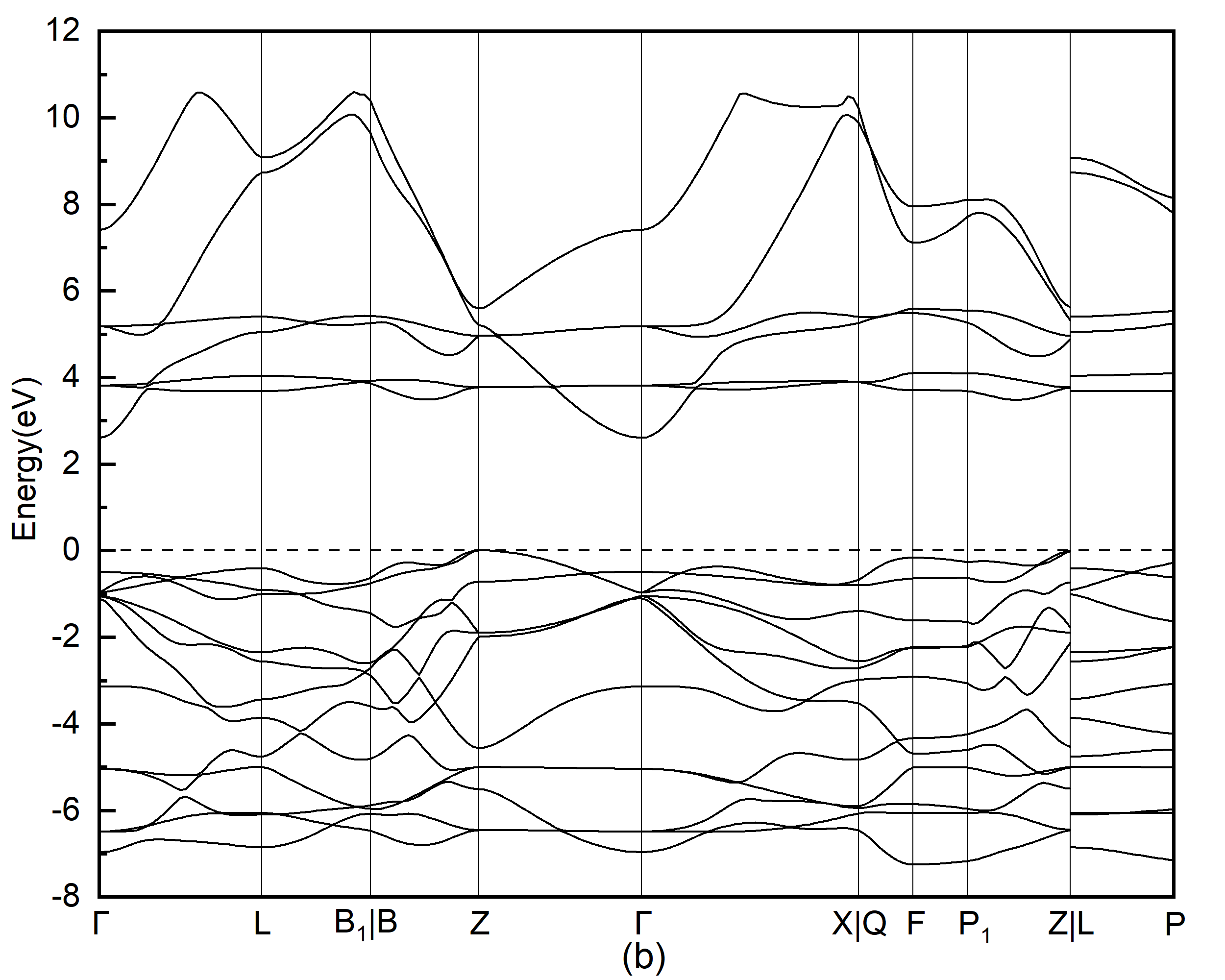}
			\includegraphics[width=\linewidth]{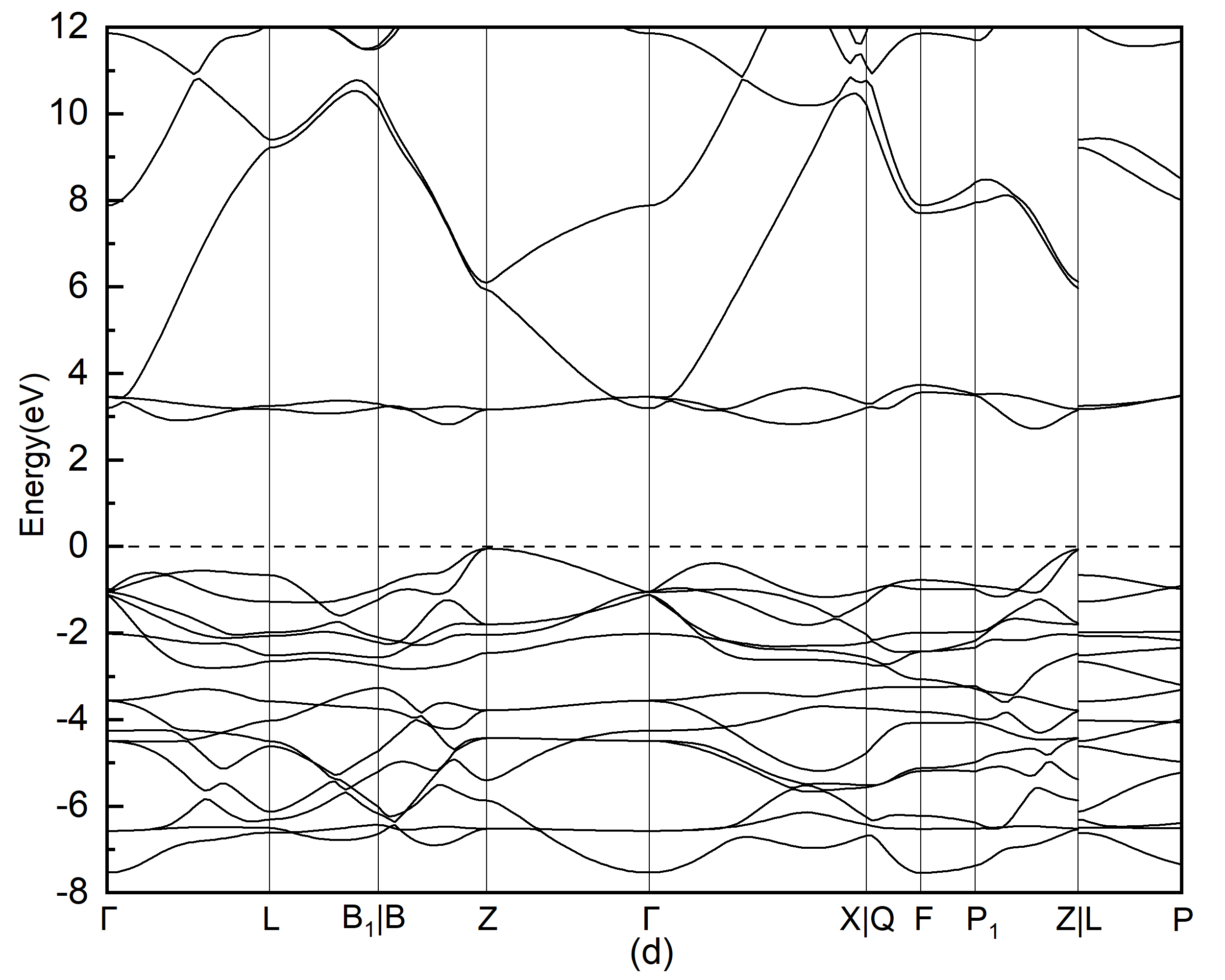}
		\end{minipage}
	}
	\caption{The ABACUS PBE+$U$ band structures of MnO, FeO, CoO, and NiO calculated with $\bar{U}$ of \SI{5.0}{\electronvolt} are depicted in panels (a), (b), (c), and (d), respectively. The energy zero is set to be the top of the occupied bands (horizontal dashed line).}
	\label{Fig.3}
\end{figure*}

 To gain more insights about the performance of our DFT+\textit{U} scheme, we calculate the projected density of states (PDOS) and band structures of MnO, FeO, CoO and NiO at a fixed effective on-site Coulomb interaction parameter $\bar{U}$ of \SI{5.0}{\electronvolt},  and the obtained results are plotted in Fig.~\ref{Fig.2} and Fig.~\ref{Fig.3}, respectively . In the type--\uppercase\expandafter{\romannumeral2} AFM unit cells, the spin polarizations of the two neighboring 
TM atoms are of the same magnitude but anti-parallel in direction. In Fig.~\ref{Fig.2},
we project the total density of states (TDOS) to the spin-up polarized TM atoms and their nearest oxygen atoms. Figure~\ref{Fig.2}(a) indicates that 3\textit{d} electrons of the projecting site Mn are nearly fully spin-up polarized, which suggests all 3\textit{d}  electrons occupy the spin up states and this result agrees with the previous work using WIEN2k, VASP and QE codes \cite{Wang2016}. As the number of $d$ electrons increases from Mn to Ni, the spin-down $d$ states 
also get populated, resulting in a decrease of the total magnetic moments (cf. Table~\ref{table2}). In all four mono-oxides, there are significant hybridizations between
the O 2\textit{p} and TM 3\textit{d} orbitals over a wide range of energy. In particular, 
the top valence states have a predominant contribution from the O 2$p$ orbitals, suggesting a charge transfer character of these TMOs. Such behaviors agree with the chemistry bond theory and previous theoretical results \cite{Anisimov1991,Cococcioni2005, Han2006, Wang2016}. 

\par We further present the calculated band structures of the four TMOs in Fig. \ref{Fig.3}, which display the typical character of strongly correlated systems. The occupied manifold of the KS states is a mixture of the O $2p$ and TM $3d$ characters. Energetically the TM 3$d$ dominating bands sit below the O 2$p$ dominating ones, due to the
fact that the strong Coulomb repulsion pushes the occupied TM 3$d$ states deeply down in energy. Furthermore, 
when going from MnO to NiO, 
a part of the low conduction bands transfers into the valence bands, while maintaining an insulating band gap of about
3 eV (cf. Fig.~\ref{Fig.3}(a)-(d)). Both the top valence bands and bottom conduction bands  show a very
small dispersion of the order of \SI{1.0}{\electronvolt}, 
which is much smaller when compared to the effective on-site Coulomb interaction energy $\bar{U}$. Theoretically, the hoping amplitude of electrons between neighboring lattice sites is proportional to the bandwidth. In the cases where the bandwidth is much smaller than the on-site Coulomb interactions, the transport process of the valence electrons gets hampered and it
is very difficult for them to hop  between neighboring sites, so than 3\textit{d} electrons become localized, leading to the insulating behavior. The narrow widths of the top valence bands agree well with this physics picture of strongly correlated systems. All these properties suggest that our DFT+\textit{U} implementation successfully captures the fundamental physical mechanism of strongly correlated TM monoxides.

\par
\begin{figure}[htb]
	\graphicspath{{figures/}}
	\centering
	\includegraphics[width=\linewidth]{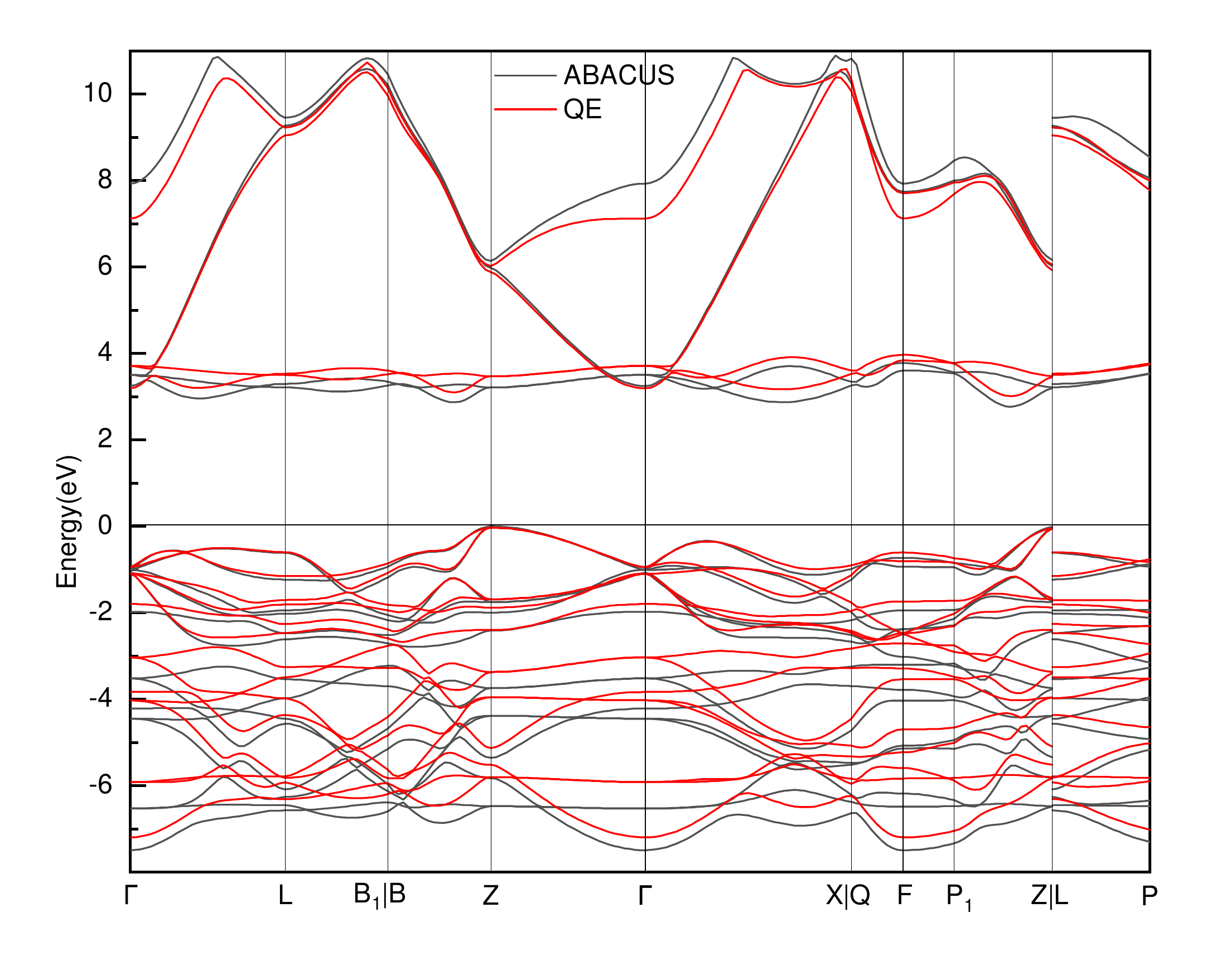}
	\caption{The PBE+$U$ band structures of NiO calculated by ABACUS and QE with $\bar{U}=$\SI{5}{\electronvolt}. The zeroes of energy is set to be the top of highest occupied bands .}
	\label{Fig.4}
\end{figure}

 As a comparison to another implementation scheme, Fig.~\ref{Fig.4} presents the PBE+$U$ band structures of NiO calculated by ABACUS and the plane-wave based code QE at $\bar{U}$ of \SI{5.0}{\electronvolt}. The DFT+\textit{U} bands of NiO given by the two codes are fairly close, with the occupied $d$ bands of ABACUS slightly lower in energy compared to the QE results. This difference may be caused by the different underlying DFT+\textit{U} schemes of the two codes. It
should be noted that the same $\bar{U}$ value can
lead to somewhat different effects within different DFT+\textit{U} schemes. 

\subsection{Validation of the force and stress calculations}\label{sec:4c}
\par To assess the validity of our DFT+$U$ force and stress implementations, we compare the results calculated via the analytical derivative
formalism in Eq.~(\ref{eq:C6}) and (\ref{eq:C9}) with those obtained by the finite-difference (FD) method, taking NiO in the conventional cell as the testing system.  
We first calculate the atomic forces for a series of structures with the Ni atom moving along the $z$-axis while other atoms staying at their original positions, and the results are reported in Table~\ref{force-finite-difference}.
As clearly shown in Table~\ref{force-finite-difference}, the forces based on analytical gradients and those determined 
by the FD method agree fairly well, with the remaining discrepancy below $2\%$.
		
		
		

		
\begin{table}[ht]
	\centering
	\caption{Benchmark force calculations for NiO. The 1st column is the $z$ coordinate of the Ni atom, and the 2nd column
	shows the energy increase for each move of the Ni atom. The 3rd and 4th columns present the FD and analytical forces (in eV/\AA) experienced by the Ni atom, respectively.
	The last column presents the relative deviations (in percentage) of the results computed by the two approaches. }     
	\begin{ruledtabular}
	\begin{tabular}{p{1.2cm}<{\centering} p{1.2cm}<{\centering} p{1.4cm}<{\centering}  p{1.3cm}<{\centering} p{1.4cm}<{\centering}}
		$z$ (\AA)  &$\Delta E$ (eV)  & $\Delta E / \Delta z$ & Force  & Deviation\\
		\colrule
		2.297820  &             &              & -2.563364    &   \\
		
		2.301998  & 0.010869    & -2.601461    & -2.637573    & 1.37\%    \\
		
		2.306175  & 0.011193    & -2.679244    & -2.712991    & 1.24\%   \\
		
		2.310353  & 0.011472    & -2.745873    & -2.789429    & 1.56\%    \\

        2.314531  & 0.011831    & -2.831828    & -2.868811    & 1.29\%    \\
		
	\end{tabular}
   \label{force-finite-difference}
	\end{ruledtabular}
\end{table}
For benchmark stress calculations, we fix the lattice constants $a$ and $b$ and vary the length of $c$ of the NiO (conventional cubic) cell, 
and the results are presented in Table \ref{stress-finite-difference}.
Comparison of the  FD and analytical stress results shows that the relative deviations are in the order of 2-3$\%$. Such level of accuracy is adequate
for relaxing the lattice structures. 
    
    
    

  

\begin{table}[ht]
  \centering
  \caption{Benchmark stress calculations for NiO. The 1st column is the length of the lattice constant $c$ of the conventional unit cell and the 2nd column shows the 
  energy change of each increase of $c$. The 3rd and 4th columns present the FD and analytical stress results (in kbar), respectively. 
  In the 3rd column, the FD stress is calculated by $\Delta E / {\epsilon \Omega}$ where $\epsilon$ is the strain component of \textit{c} direction, i.e., $\epsilon_{cc}$ 
  and $\Omega$ is the cell volume.}
  \begin{ruledtabular}
  \begin{tabular}{p{1.2cm}<{\centering} p{1.2cm}<{\centering} p{1.5cm}<{\centering} p{1.5cm}<{\centering} p{1.4cm}<{\centering}}
    $c$ (Bohr)  &$\Delta E$ (eV)  & $\Delta E / {\epsilon \Omega}$ & Stress  & Deviations\\
    \colrule
    7.895000  &             &               & -47.484882   &   \\
    
    7.902895  & 0.002266    & -52.926873    & -51.879179   & -2.02\%    \\
    
    7.910790  & 0.002464    & -57.497658    & -56.164020   & -2.37\%   \\
    
    7.918685  & 0.002666    & -62.141370    & -60.327858   & -3.01\%    \\ 

    7.926580  & 0.002842    & -66.188954    & -64.360639   & -2.84\%    \\
  
  \end{tabular} \label{stress-finite-difference}
  \end{ruledtabular}
\end{table}


\subsection{DFT+$U$+SOC band structure}\label{sec:4d}
\par
\begin{figure}[htb]
	\graphicspath{{figures/}}
	\centering
	\begin{minipage}[htb]{0.8\linewidth}
		\centering
		\includegraphics[width=\linewidth]{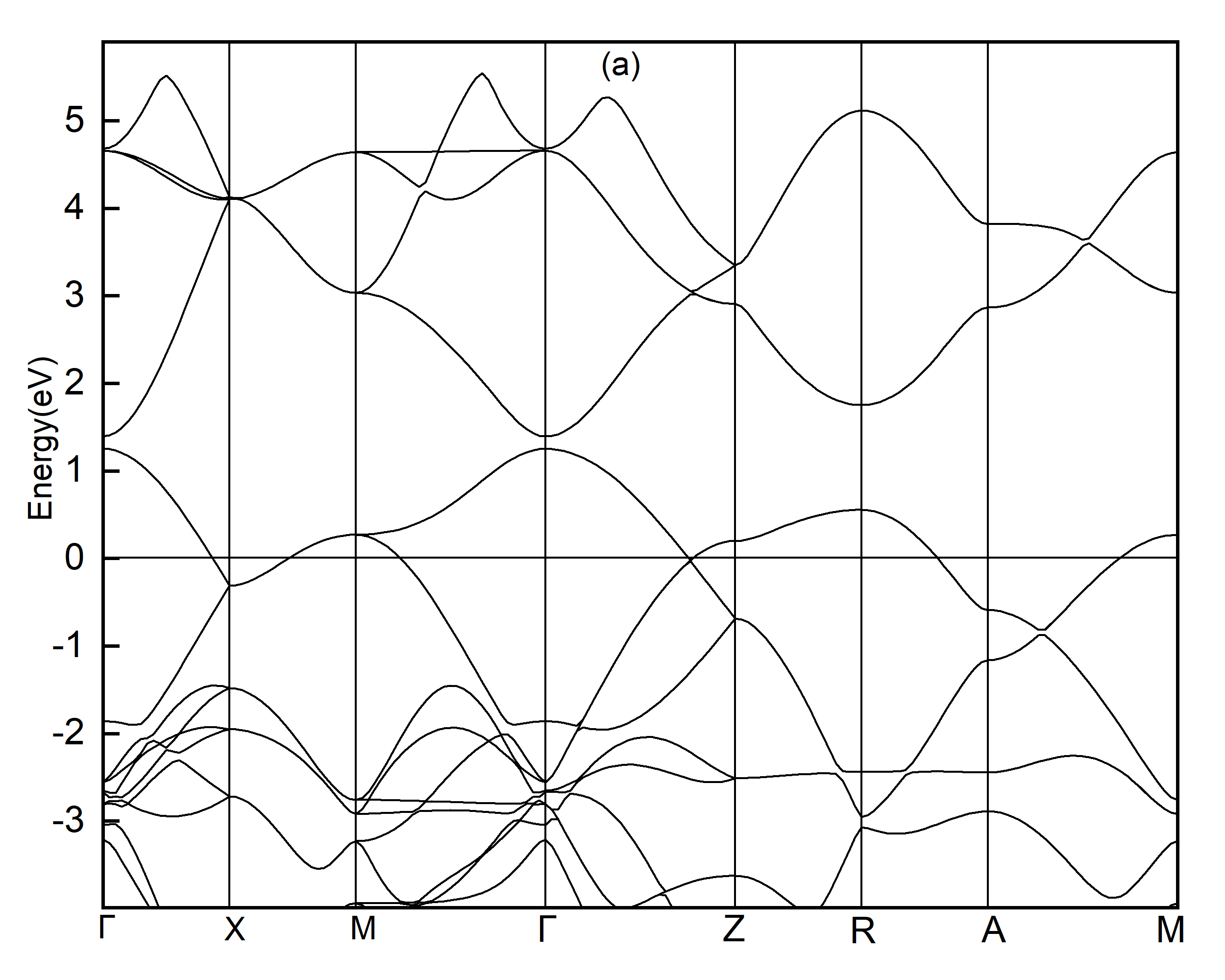}
		\includegraphics[width=\linewidth]{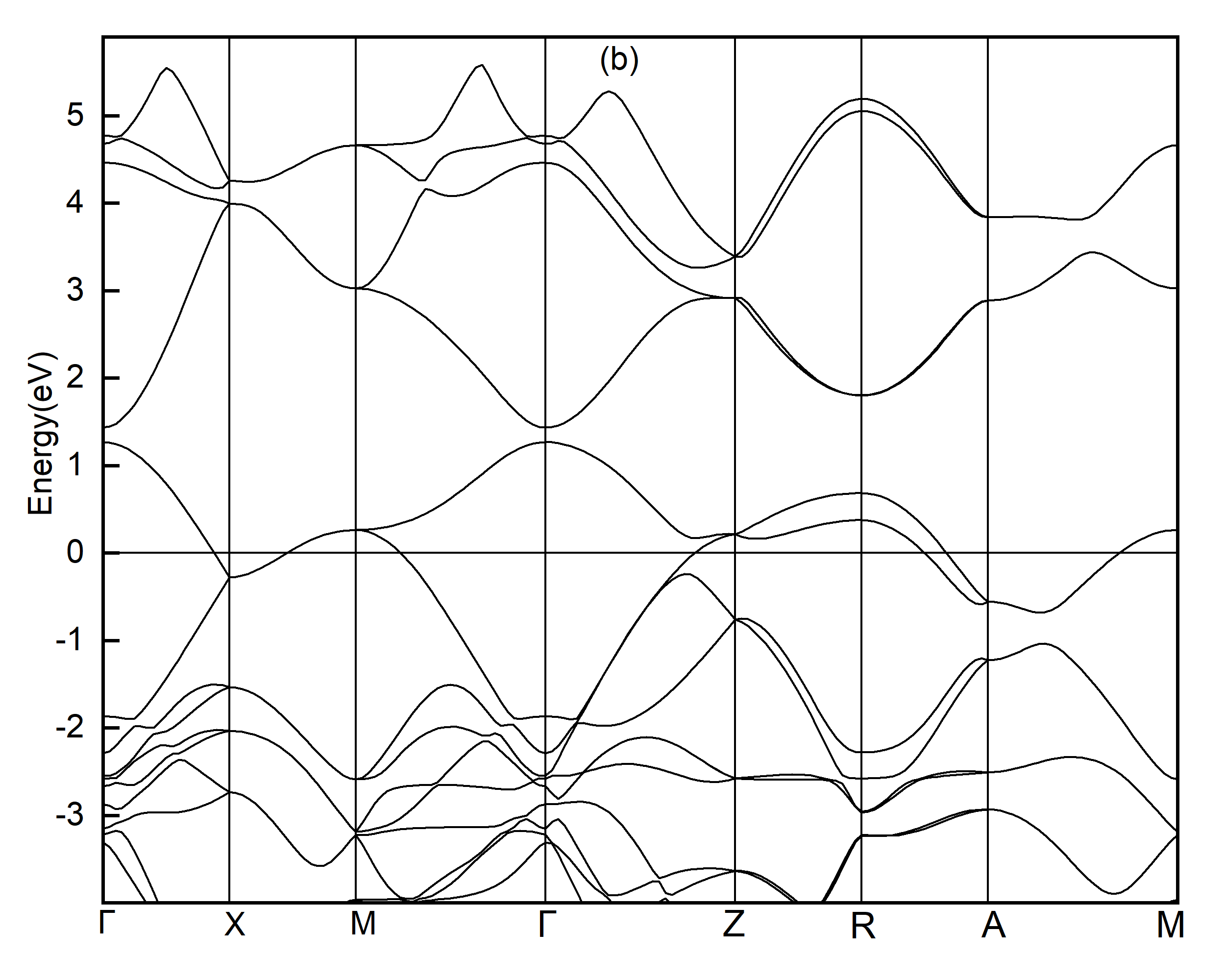}
		\includegraphics[width=\linewidth]{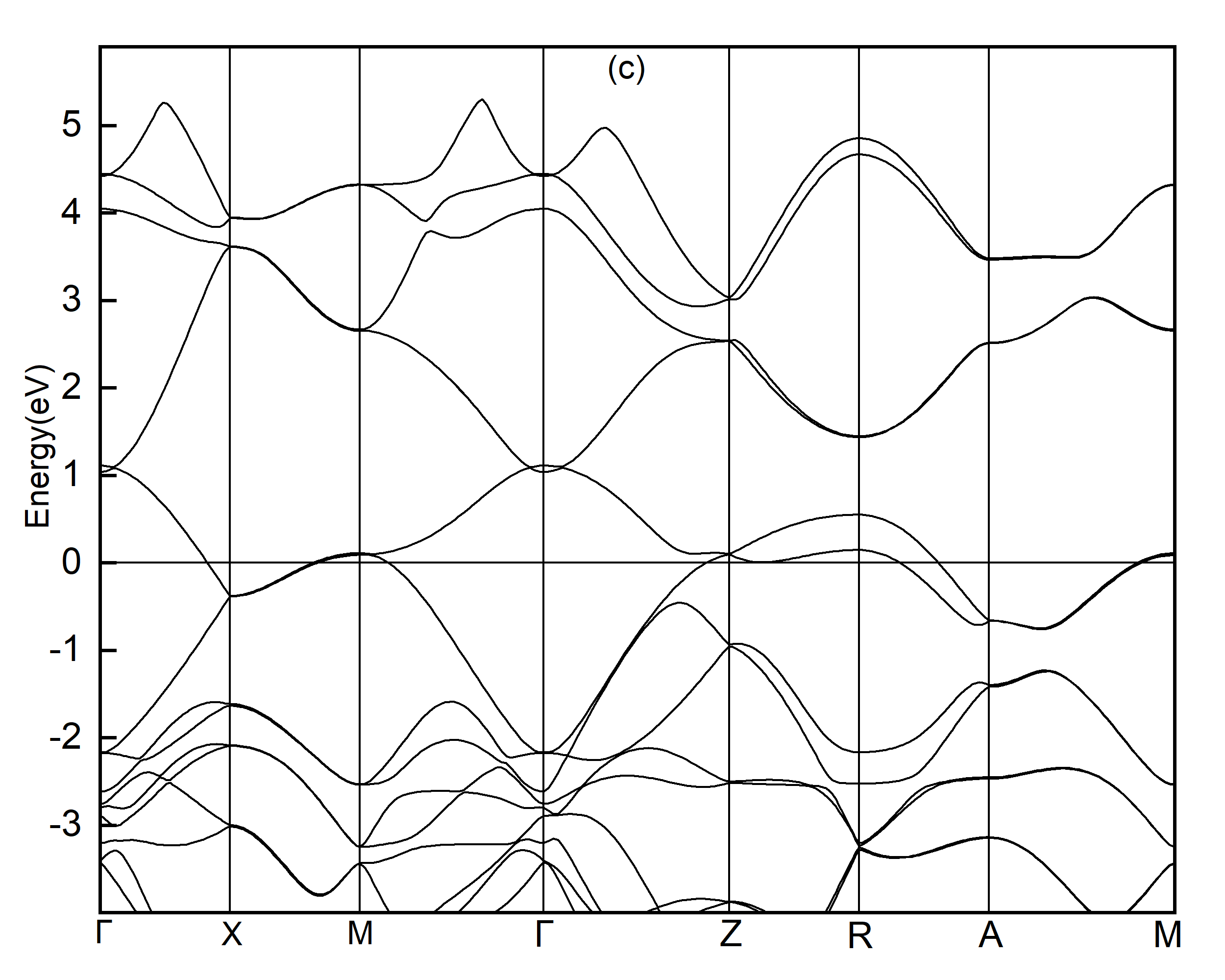}
	\end{minipage}
	\caption{Band structures of IrO$_2$ calculated by (a) ABACUS DFT+\textit{U}, (b)ABACUS DFT+\textit{U}+SOC, and (c) VASP DFT+\textit{U}+SOC schemes.}
	\label{Fig.5}
\end{figure}

 The strength of SOC scales as $O(Z^4)$, where $Z$ is the atomic number. The atomic number of Iridium (Ir) is 77, which is nearly three times larger than that of Mn (atomic number 25). Thus theoretically the SOC effect of Ir-based compounds is much stronger than 3\textit{d} TM materials. In this subsection, we examine the band structure of IrO$_2$ to assess the performance of our DFT+\textit{U}+SOC implementation. We consider the AFM structure with the magnetic momenta of two neighboring Ir atoms in the conventional cell  antiparallel along the $z$-axis. The DFT+\textit{U} correction is applied to the 5\textit{d} electrons of Ir atoms with $\bar{U}=\SI{2.0}{\electronvolt}$. The 
calculated PBE+$U$ and PBE+$U$+SOC band structures are presented in Fig.~\ref{Fig.5}(a) and Fig.~\ref{Fig.5}(b), respectively.

\par The PDOS analysis (not shown) suggests that the main components of the bands located in the energy window included in Fig.~\ref{Fig.5} are 5\textit{d} electrons of 
the Ir atom. 
Comparing Fig.~\ref{Fig.5}(a) and \ref{Fig.5}(b), the originally degenerate bands along the $\bfk$-path from Z to A are split into subbands after the introduction of SOC, 
which is the typical effects of SOC on heavy elements.

\par For comparison, in Fig.~\ref{Fig.5}(c) the PBE+$U$+SOC band structure calculated by VASP are also presented. As can be seen, ABACUS and VASP give similar band structures 
for IrO$_2$. The small difference in details between ABACUS (\ref{Fig.5}(b)) and VASP (\ref{Fig.5}(c)) results may stem  from different basis sets, different
pseudopotentials and the different DFT+\textit{U} and SOC schemes. In the latter case,  in VASP the SOC is included in a non-self-consistent way based on a perturbation approach,
whereas in ABACUS, the SOC is treated in a self-consistent way. In addition, the results in Fig.~\ref{Fig.5} are in good agreement with previous theoretical work using all-electron full-potential DFT+\textit{U}+SOC and LDA+DMFT method \cite{Panda2014}.  

\subsection{Yukawa-potential based $U$ and $J$ parameters}\label{sec:4e}
In the above discussions, we focused on the validation of our DFT+$U$ implementation whereby $U$ and $J$ are treated as 
given parameters. 
To apply DFT+$U$ to novel materials where no reference results are available, it is crucial to be able to determine these parameters from first principles calculations.
As mentioned above, in the literature, different theoretical schemes have been developed to determine these parameters 
\cite{Gunnarsson1989,Anisimov1991,Aryasetiawan2004,Miyake2008,Miyake2009,Sakuma2013,Cococcioni2005}. In this subsection, we check how the scheme based on the Yukawa potential, 
as described in Sec.~\ref{sec:Yukawa}, works for determining $U$, $J$ parameters.

As discussed in Sec.~\ref{sec:Yukawa}, for a given set of NAO basis functions, the calculated interaction parameters based on Eq.~(\ref{E1}) depend solely on the screening parameter
$\lambda$. Here, we first examine how the Coulomb interaction parameter $U$ and the exchange parameter $J$ change with the $\lambda$ value for the local $d$/$f$-type orbitals 
We take three TMOs, i.e., MnO, FeO, and NiO as test examples, and use NAO DZP basis sets in the calculations. For the TMs Mn, Fe, and Ni, the DZP basis set
contains two $d$ functions and one $f$ function. The Yukawa potential scheme allows one to readily compute the $U$, $J$ parameters for all these orbital functions.

\par Table~\ref{table3} presents the calculated \textit{U} and \textit{J} parameters for the two $d$ and one $f$ orbitals with $\lambda$ varying from 0.80 to \SI{1.08} {Bohr^{-1}}. First, one can see
that the $U$, $J$ values for the first $d$ orbital are one order of magnitude larger than those for the second one. This is because the first $d$ orbital is localized and
has no node, whereas the second orbital, designed to be orthogonalized to the first one, is much more extended and has a node (see the Appendix for further details). 
Thus, it is not surprising that the $U$, $J$ corresponding to
the second $d$ orbital are much smaller than those of the first one. According to Table~\ref{table3}, one can see
that the second $d$ orbital can be safely excluded from the correlated subspace in the DFT+$U$ calculations.
Furthermore, for these TM compounds,
the $f$ functions are ``polarized orbitals" and represent orbital space that is well separated from the correlated subspace around the Fermi level, and 
consequently require no special treatment. 
Consistent with this, the $U$, $J$ parameters for the $f$ orbital are also significantly smaller than those of the first $d$ orbital. 

Table~\ref{table3} shows that the obtained $U$, $J$ values steadily decrease upon increasing the $\lambda$ value, which is easy to understand since a larger $\lambda$
means a stronger screening of the Coulomb interaction. Focusing on the first $d$ orbital ($d1$ in Table~\ref{table3}) and
comparing to the results given by constrained DFT \cite{Jiang2010}, constrained RPA \cite{Sakuma2013} and local screened Coulomb correction (LSCC) scheme \cite{Wang2019}, we find that a screening parameter $\lambda$ around \SI{1.00} {Bohr^{-1}} yields $U$, $J$ values that are fairly close to those reported in the literature. Remarkably, it seems that $\lambda \approx \SI{1} {Bohr^{-1}}$ is a sensible choice that works well for all three TMOs. Therefore, at least the TM compounds, our DFT+$U$ scheme together with the 
Yukawa-potential scheme for determining $U$, $J$ parameters can be viewed as a semi-empirical approach that has predictive power; namely, the only input parameter
is fixed and does not vary for different materials.


\begin{table*}[htbp]
	\centering
	\caption{On-site Coulomb energy $U$ and exchange energy $J$ (in eV) as a function of screening parameter $\lambda$ (in Bohr$^{-1}$) for MnO, FeO and NiO in DZP basis.Here \textit{d} and \textit{f} are angular momentum indexes while the number behind them are multiplicity indexes $\zeta$.}
	\resizebox{\textwidth}{!}{
	\begin{ruledtabular}
	\begin{tabular}{c cc p{0.05cm}<{\centering} cc p{0.05cm}<{\centering} cc c cc p{0.05cm}<{\centering} cc p{0.05cm}<{\centering} cc c cc p{0.05cm}<{\centering} cc p{0.05cm}<{\centering} cc}

		 \multirow{3}*{$\lambda$}    &\multicolumn{8}{c}{MnO}  &    & \multicolumn{8}{c}{FeO} &   & \multicolumn{8}{c}{NiO} \\
		   \cline{2-9} \cline{11-18} \cline{20-27}
		   &\multicolumn{2}{c}{\textit{d}1} & &\multicolumn{2}{c}{\textit{d}2}& &\multicolumn{2}{c}{\textit{f}1}&  &\multicolumn{2}{c}{\textit{d}1}& &\multicolumn{2}{c}{\textit{d}2}& &\multicolumn{2}{c}{\textit{f}1}&   &\multicolumn{2}{c}{\textit{d}1}& &\multicolumn{2}{c}{\textit{d}2}& &\multicolumn{2}{c}{\textit{f}1} \\
		   \cline{2-3} \cline{5-6} \cline{8-9} \cline{11-12} \cline{14-15} \cline{17-18} \cline{20-21} \cline{23-24} \cline{26-27}
		   &$U$&$J$ &  &$U$&$J$& &$U$&$J$& 
		   &$U$&$J$& &$U$&$J$& &$U$&$J$& 
		   &$U$&$J$& &$U$&$J$& &$U$&$J$ \\
		\colrule
		0.80  &6.62&1.00& &0.62&0.20& &2.93 &0.49&  
		          &7.41&1.08& &0.57&0.20& &2.80 &0.48&
		          &8.04&1.20& &0.49&0.20& &2.29 &0.44 \\
		          		
		0.84  &6.35&1.00& &0.57&0.19& &2.77 &0.48&
		          &7.12&1.07& &0.53&0.19& &2.65 &0.48&
		          &7.74&1.19& &0.45&0.19& &2.16 &0.44  \\
		
		0.88  &6.09&0.99& &0.53&0.18& &2.62 &0.47&  
		          &6.84&1.06& &0.49&0.18& &2.51 &0.47&
		          &7.44&1.18& &0.42&0.18& &2.04 &0.43 \\
		
		0.92  &5.85&0.98& &0.50&0.18& &2.48 &0.47&
		          &6.58&1.05& &0.46&0.18& &2.38 &0.46&
		          &7.16&1.17& &0.39&0.18& &1.93 &0.42   \\
		
		0.96  &5.62&0.97& &0.47&0.17& &2.36 &0.46&
		         &6.33&1.04& &0.43&0.17& &2.26 &0.46&
		         &6.90&1.16& &0.36&0.17& &1.82 &0.42    \\
		
		1.00  &5.40&0.96& &0.44&0.16& &2.24 &0.45&
		         &6.10&1.03& &0.40&0.16& &2.14 &0.45& 
		         &6.65&1.15& &0.34&0.16& &1.73 &0.41   \\
		         
		 1.04  &5.20&0.95& &0.41&0.16& &2.13 &0.45&
		 &5.87&1.03& &0.38&0.16& &2.04 &0.44& 
		 &6.41&1.15& &0.32&0.16& &1.64 &0.40   \\
		 
		 1.08  &5.00&0.94& &0.39&0.15& &2.03 &0.44&
		 &5.66&1.02& &0.35&0.15& &1.94 &0.44& 
		 &6.18&1.14& &0.30&0.15& &1.56 &0.40   \\
		\midrule      
        cDFT$^a$ &\multicolumn{8}{c}{$U$=4.7, $J$=0.8 }  &
		                   & \multicolumn{8}{c}{$U$=4.8, $J$=0.9} & 
		                   & \multicolumn{8}{c}{$U$=5.2, $J$=0.9 } \\
		
		cRPA$^b$ &\multicolumn{8}{c}{$U$=5.5, $J$=0.6}  &
		& \multicolumn{8}{c}{$U$=5.7, $J$=0.7} & 
		& \multicolumn{8}{c}{$U$=6.6, $J$=0.7} \\
		
	\end{tabular}
    \label{table3}
	\end{ruledtabular}
}
	\begin{tablenotes}
		\item[1] $^\text{a}$The results of constrained DFT implemented in LAPW framework from Ref.\cite{Jiang2010}
		\item[3] $^\text{b}$The results of constrained RPA in maximally localized Wannier functions from Ref.\cite{Sakuma2013}
	\end{tablenotes}
\end{table*}

Although a $\lambda$=\SI{1.00}{Bohr^{-1}} seems to work well for all TMOs, we don't expect it to work in general cases, 
because a fixed $\lambda$ value means that the obtained $U$, $J$
values only depend on the atomic species and the chosen NAO basis sets, but not on the chemical environment. 
For a generally applicable scheme, $\lambda$ should reflect the chemical environment of the 
system. The averaged $\bar{\lambda}$ introduced in Eq.~(\ref{E7}) depends on the electron density of the system via the Thomas-Fermi screening model, and thus accounts for the chemical environment in a natural way. In Ref.~\cite{Wang2019},
it has been shown that such a scheme (termed as LSCC there) yields rather good $U$, $J$ values within the LAPW framework.
Note that, within such a scheme, the averaged screening parameter $\bar{\lambda}$ varies during the self-consistent 
iterations and hence so do the $U$, $J$ values, until the convergence is reached.
We also implemented the LSCC scheme in ABACUS, and the self-consistently determined $\bar{\lambda}$ values
for MnO, FeO, CoO, and NiO are 1.581,1.627,1.649 and 1.677 Bohr$^{-1}$, respectively. 
However, the $\bar{\lambda}$ parameters lead to too short screening lengths for our NAO basis sets and 
the calculated $U$, $J$ values are too small. 
In practice, we find that introducing a scaling factor of 0.625 can reduce the $\bar{\lambda}$ value to a range
(around 1.0 Bohr$^{-1}$) that yields physically reasonable $U$, $J$ values. The reason that an additional scaling
factor is needed here, compared to the original LAPW-based LSCC implementation, is that the local NAOs used here is
more extended than the local orbitals in the LAPW framework, which are restricted within the muffin-tin sphere. 
Hence the computed interaction parameters within the NAO framework will be smaller than the LAPW case 
if the same screening parameter is used.

In Table~\ref{table4},
we present the energy differences between the AFM and ferromagnetic (FM) states of four materials, as computed by PBE and
PBE+$U$. The experimental results and the results reported in Ref.~\cite{Wang2019} are included for comparison. 
The PBE+$U$ with a scaled $\bar{\lambda}$ parameter, obtained from the Thomas Fermi model, yields results that 
show satisfactory agreement with experiment and the LSCC results. The accuracy of the results is a factor of two better 
than that of PBE. 

\begin{table}[htb]
	\centering
	\caption{ The energy differences (in meV) between the AFM and FM states of four materials calculated 
	by PBE and PBE+$U$ with $U$ and $J$
	determined with a scaled screening parameter (see the text), and compared to the experimental values and
	the calculated results reported in Ref.~\cite{Wang2019}. In LSCC, the $U$, $J$ parameters are obtained
	without rescaling the $\bar{\lambda}$ parameter whereas in our work, a scaling factor of 0.625 is used.}
	\begin{ruledtabular}
	\begin{tabular}{p{0.3cm}<{\centering} p{1.4cm}<{\centering} p{1.2cm}<{\centering} p{1.4cm}<{\centering} p{1.4cm}<{\centering} p{1.2cm}<{\centering}}
	  & PBE  & PBE \cite{Wang2019} &LSCC \cite{Wang2019} & PBE+$U$  & Expt. \\
	  & (this work) &  &   & (this work) &  \\
		\colrule
		MnO  & -157.3     & -152  & -99.8  & -84.5    & -62\cite{Pepy1974}   \\
		
		NiO  & -258.2      & -261  & -107   & -145.4    & -112\cite{Shanker1973}  \\
		
		MnF$_2$  & -64.6      & -60.5  & -28.3    & -22.2    & -15.2\cite{Feng2004} \\
		
		NiF$_2$  & -85.9     & -69.6   & -20.1  & -31.7    & -13.8\cite{Feng2004}  \\
	\end{tabular}
    \label{table4}
	\end{ruledtabular}
\end{table}

\section{Summary}
\label{sec:summary}
\par We present a detailed formulation of the DFT+\textit{U} method within the framework of
NAO basis set. The key in this formulation is to use a symmetrized Mulliken charge projector, constructed in terms of
the most localized $d$ or $f$ orbital basis functions, to project a correlated local subspace out of the full 
KS orbital space. We implemented such a scheme within the ABACUS code package, and our implementation 
allows not only self-consistent electronic structure calculations with or without including the SOC effect, but also 
enables force and stress calculations.
The efficacy of our formalism and implementation has been demonstrated for the prototypical TMOs and IrO$_2$.
Furthermore, we tested the scheme for computing the $U$, $J$ parameters based on a screened Yukawa potential,
and found that, while a fixed screening parameter works for all TMOs, determining such a parameter from the
electron density via the Thomas-Fermi model leads to an underestimation of the $U$, $J$ values. However, this issue can be
fixed by introducing a universal rescaling parameter to increase the screening length. We believe that the experience
gained in the present work will be very helpful for developing Hubbard-type local correction scheme within the atomic-orbital
basis set framework.

\acknowledgements
This work is supported by National Natural Science Foundation of China (Grant Nos. 12134012, 11874335, 21873005)
and the Strategic Priority Research Program of Chinese Academy of Sciences (Grant No. XDPB25). 
We thank Dr. Wenshuai Zhang for generating the optimized NAO basis sets used in the present work,
available at the official ABACUS website \cite{abacusweb}.  Parts of the calculations are done 
on the supercomputing system in the Supercomputing Center of USTC.

\appendix*
\section{The influence of the choice of local orbitals on DFT+\textit{U}}
\label{sec:appendix}
\par In this appendix, we investigate two issues about employing the local atomic orbitals to construct the projector within 
the NAO basis set framework, which have not been elaborated in the main text. The first issue, which has already been mentioned in Sec~\ref{sec:4e}, is whether all basis orbitals belonging to characteristic \textit{d}/\textit{f} angular moment channel 
need to be included in the DFT+\textit{U} correction.  The second issue is the influence of the shape of the local orbitals
on the DFT+\textit{U} results.

\begin{table}[htb]
	\centering
	\caption{ Partial occupation numbers (sum of spin-up and spin-down channels) of Mn \textit{d} orbitals 
	of MnO as calculated by PBE. Results obtained using DZP (two $d$ orbitals) and TZDP (three $d$ orbitals) are presented.}
	\begin{ruledtabular}
		\begin{tabular}{p{1.0cm}<{\centering} p{1.0cm}<{\centering} p{1.0cm}<{\centering} p{1.0cm}<{\centering} p{1.0cm}<{\centering} p{1.0cm}<{\centering} p{1.0cm}<{\centering}}
			
			&   & $d_{xy}$ & $d_{xz}$ & $d_{z^2}$  & $d_{yz}$ & $d_{x^2-y^2}$  \\
			\colrule
			 \multirow{2}{*}{DZP} & zeta=1  & 1.372  & 1.372  & 0.773  & 1.372 & 0.773   \\
			 & zeta=2   & 0.012  & 0.012  & 0.007  & 0.012 & 0.007   \\
			\colrule	
			 \multirow{3}{*}{TZDP} & zeta=1     & 1.355  & 1.343  & 0.728   & 1.474 & 0.728   \\
			& zeta=2     & 0.030  & 0.032  & 0.028   & 0.002 &0.032   \\
			& zeta=3     & 0.002  & 0.002  & -0.006  & 0.002 & -0.006   \\
			
		\end{tabular}
		\label{table5}
	\end{ruledtabular}
\end{table}

\par
\begin{figure*}[htb]
	\graphicspath{{figures/}}
	\centering
	\subfigure{
		\centering
		\begin{minipage}[htb]{0.4\linewidth}
			\centering
			\includegraphics[width=\linewidth]{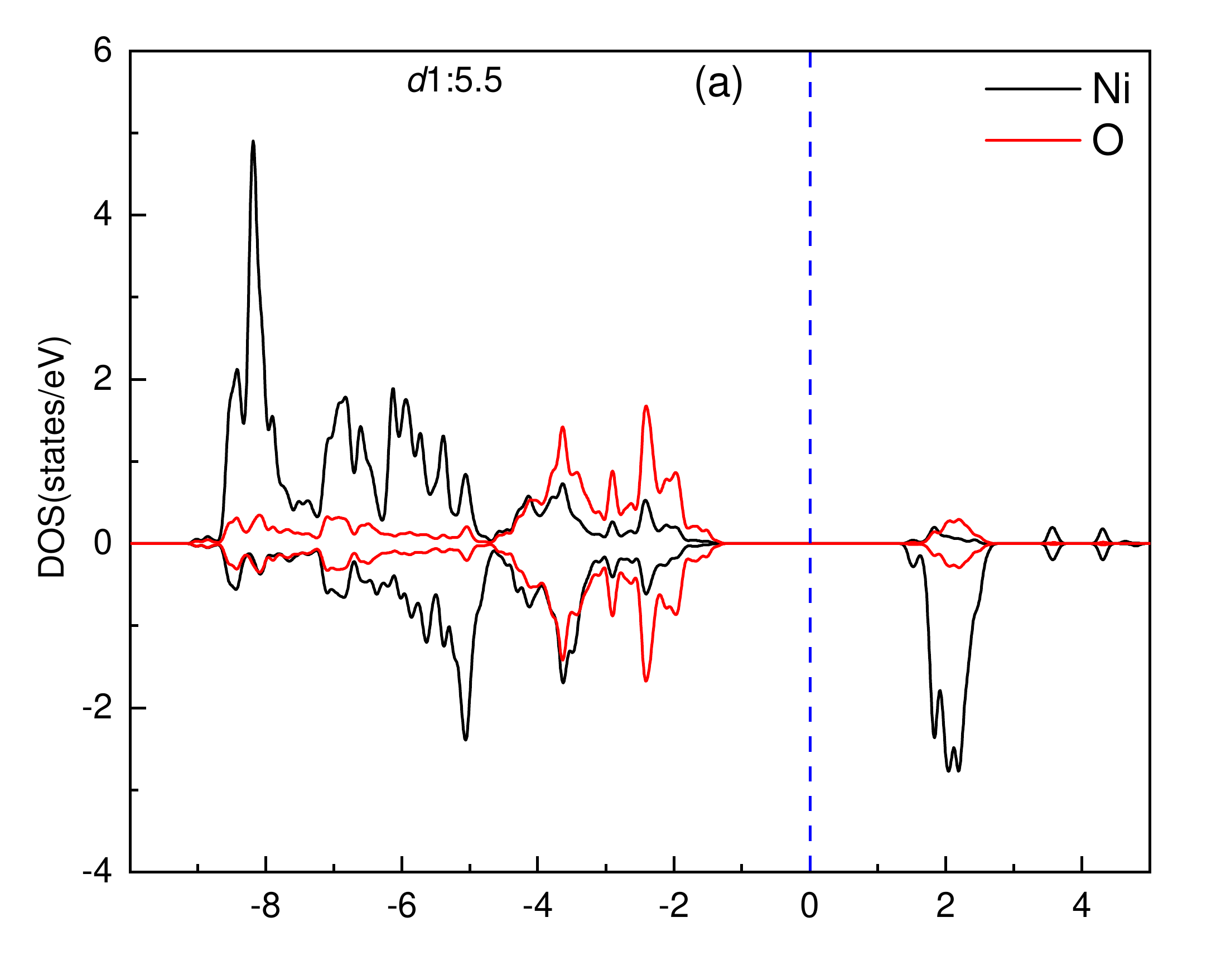}
			\includegraphics[width=\linewidth]{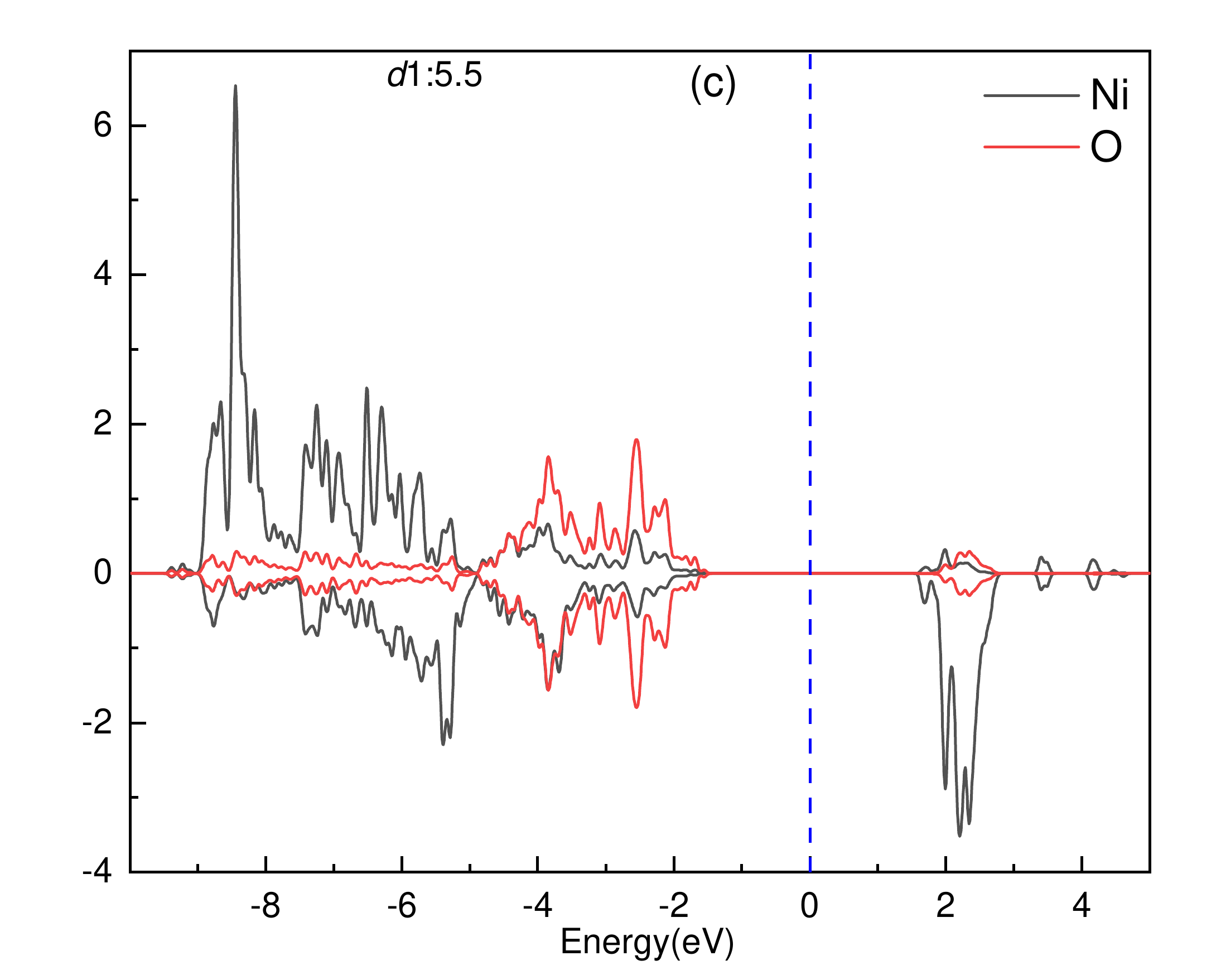}
		\end{minipage}
	}
	\hspace{5mm}
	\subfigure{
		\centering
		\begin{minipage}[htb]{0.4\linewidth}
			\centering
			\includegraphics[width=\linewidth]{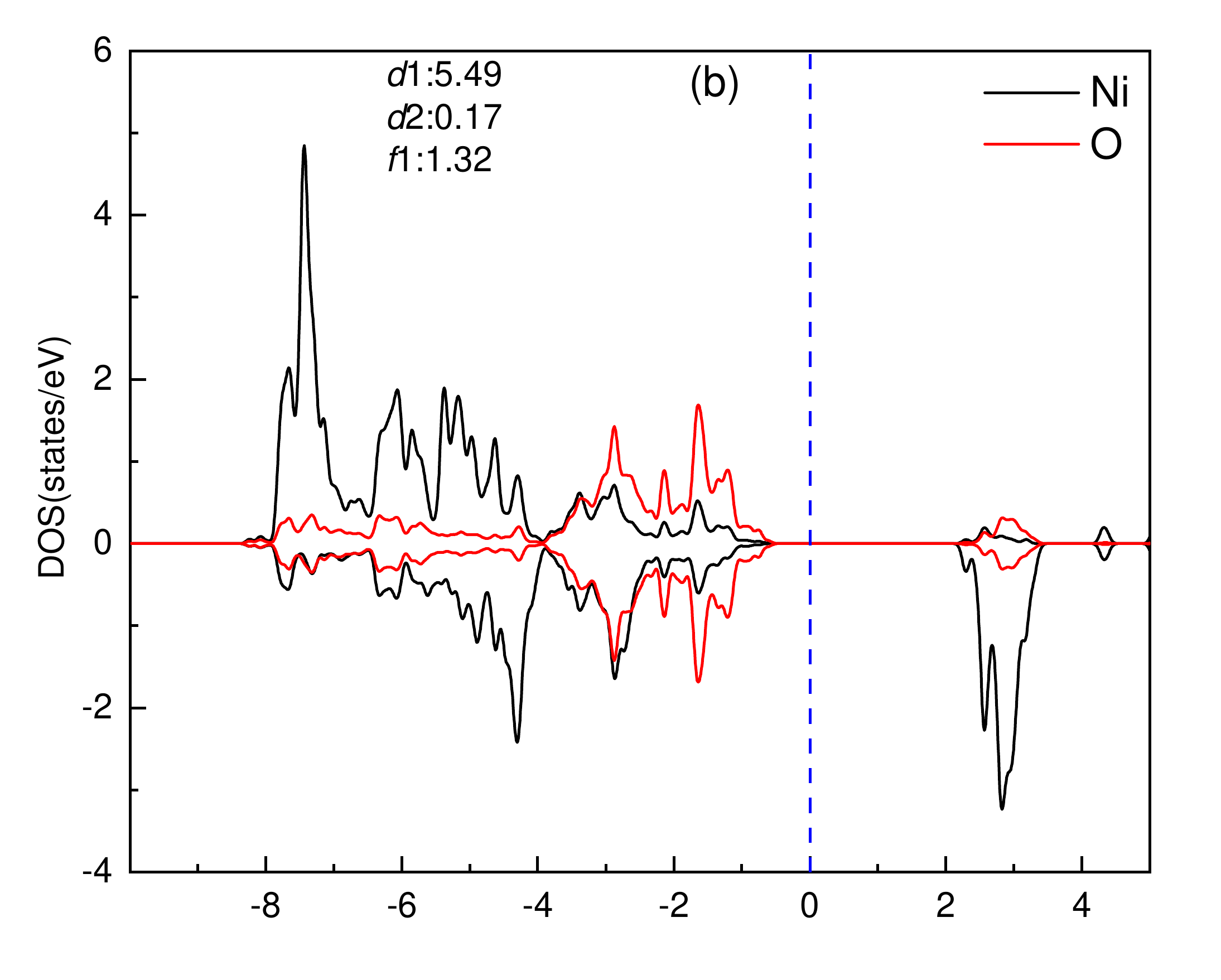}
			\includegraphics[width=\linewidth]{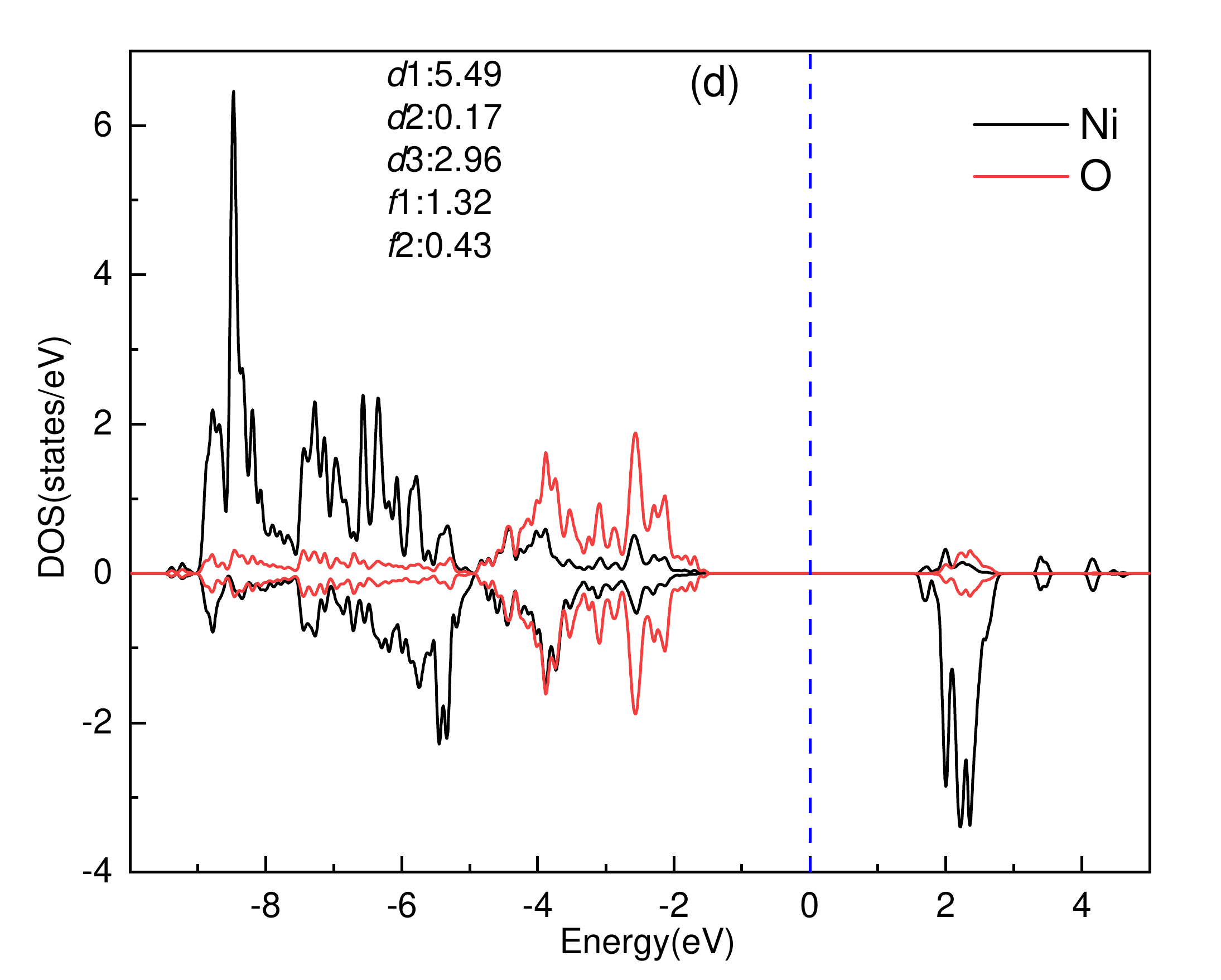}
		\end{minipage}
	}
	\caption{Projected DOS of NiO where different on-site Coulomb interactions are employed to different \textit{d}/\textit{f} orbital:(a) and (b) DZP basis, (c) and (d) TZDP basis. The upper and lower panels correspond to up-spin and down-spin. The zero of energy is set to be Fermi level (vertical dashed line).}
	\label{Fig.6}
\end{figure*}

 To address the first question, we perform a partial occupation analysis of the $d$ orbitals, taking MnO
as an example. Table \ref{table5} presents the local occupation numbers of all Mn \textit{d} orbitals of MnO, as given by Eq.~(\ref{eq:B6}). For completeness both DZP and TZDP basis sets have been used,
and in the latter case there are three $d$ orbitals for Mn. From Table~\ref{table5}, it can be clearly seen that local occupation numbers of the second or third \textit{d} orbital are significantly smaller than the first \textit{d} one. This means that the second or third $d$ orbitals contribute little to the top valence states, and the majority of the correlated local 
subspace is described by the first $d$ orbital.  From the point of view of real-space locality, the first (and innermost) \textit{d} orbital, i.e. zeta 1 in Table~\ref{table5}, is most localized and has no node. The other \textit{d} orbitals
are designed to be orthogonal to the first one, so that they have nodes and are more delocalized (cf. Fig.~\ref{Fig.6}). Such behavior is also reflected in two-electron integrals of the screened Coulomb potential.  As is
shown in Table~\ref{table4}, the on-site Coulomb interaction and exchange energies of the second \textit{d} orbital are
smaller by one order of magnitude compared to the first \textit{d} one. Both the orbital occupation and interaction 
parameter analyses suggest that
we can most likely neglect the on-site Coulomb correction to the second or higher \textit{d} orbital. 

\par Despite the observation that most probably one only needs to apply the Hubbard $U$ correction to 
the most localized orbital for multi-zeta NAO basis sets, we nevertheless also checked what if the 
$U$ correction is added to all $d$ orbitals. 
In Fig.~\ref{Fig.6}, the PDOS of NiO by using DZP and TZDP basis set on the spin-up polarized nickel atoms and their nearest oxygen atoms are shown. Figure.~\ref{Fig.6}(a) and \ref{Fig.6}(c) are the results of standard DFT+\textit{U}, where $\bar{U}$ is set to be 5.5 eV and the on-site Coulomb interaction correction is only applied to the first \textit{d} orbitals. 
For comparison, We then apply on-site Coulomb interaction corrections to all \textit{d} and polarization \textit{f} orbitals where the corresponding $\bar{U}$ value are determined by the Yukawa potential scheme as described in Sec.~\ref{sec:Yukawa} with screened parameter $\lambda$ fixed at \SI{1.00}{Bohr^{-1}}. The obtained results are depicted in Fig.~\ref{Fig.6}(b) and Fig.~\ref{Fig.6}(d) for the DZP and TZDP basis sets, respectively. In Fig.~\ref{Fig.6} the label \textit{d}/\textit{f}-\textit{i} means the \textit{i}-th \textit{d}/\textit{f} orbital and the values behind them are the corresponding $\bar{U}$ values.
As clearly shown in Fig.~\ref{Fig.6}, the PDOS results undergo little changes for both DZP and TZDP basis sets if the $U$ corrections
are added to all \textit{d} and polarization \textit{f} orbitals. This result validates our DFT+$U$ projection scheme 
that only the first, most localized correlated orbital needs to be included in the construction of the projector. 

\begin{figure}[h]
	\graphicspath{{figures/}}
	\centering
	\begin{minipage}[htb]{0.6\linewidth}
		\centering
		\includegraphics[width=\linewidth]{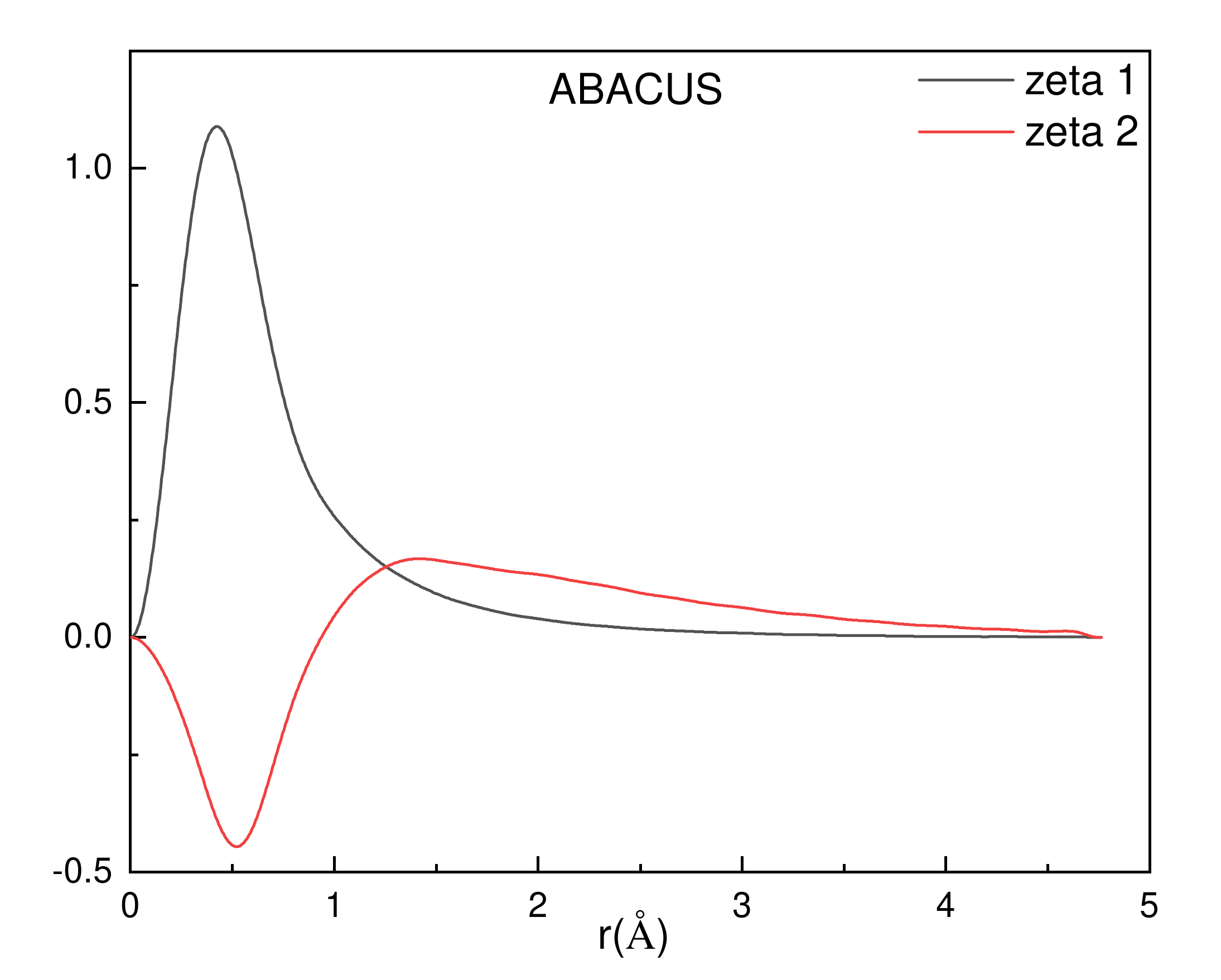}
		\includegraphics[width=\linewidth]{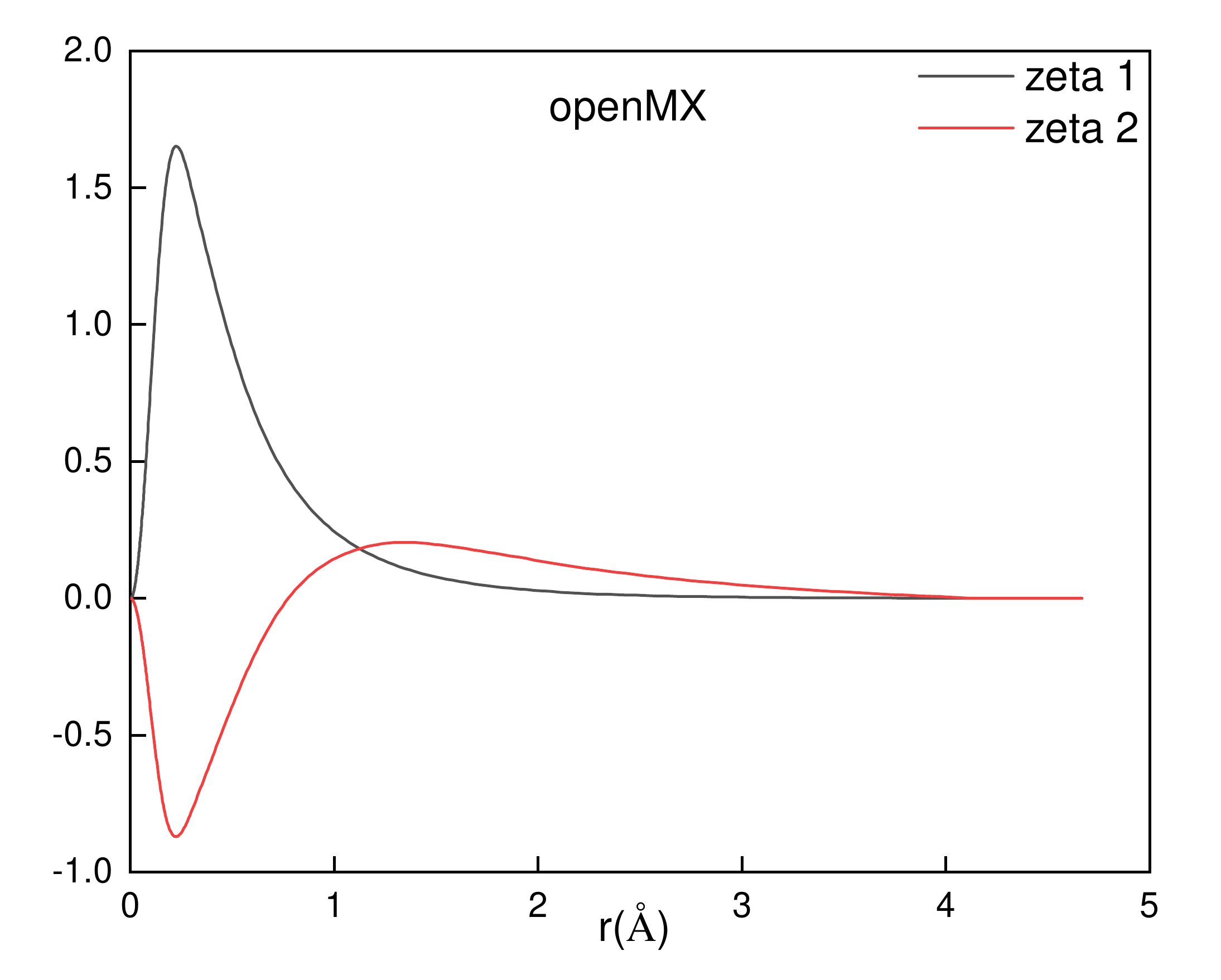}
	\end{minipage}
	\caption{Radial functions of 3d basis of Mn atom in ABACUS}
	\label{Fig.7}
\end{figure}

\par Finally, we briefly discuss the possible influence of the shape of the local atomic orbitals used in the projector 
construction on the DFT+\textit{U} calculations. From the discussion in Sec.~\ref{sec:general_formalism},
one can see that the KS wavefunctions and the local orbitals control the local occupation matrix and hence directly affect the results of DFT+\textit{U} calculations. 
Since all properly worked DFT codes must yield nearly 
the same wavefunctions in real space, provided that the same pseudopotential is used, the DFT+\textit{U} results 
mainly depend on the choice of local correlated orbitals. In Fig.~\ref{Fig.7}, we plot the radial functions of Mn $d$ orbitals
of the DZP basis sets, employed in ABACUS (upper panel) and OpenMX (lower panel) calculations. 
For both codes, the first Mn \textit{d} orbitals (the nodeless ones) are localized within 1.5 {\AA} around the
nucleus, with a sharp peak positioned around $r=\SI{0.5}{\angstrom}$. The local occupation numbers given by the Mulliken charge projector in OpenMX are 1.447 for $t_{2g}$ orbitals and 0.412 for $e_g$ orbitals, in comparison to 1.372 and 0.773 as given
by ABACUS (cf. Table \ref{table5}). Such difference is a manifestation of the difference in the radial shape of 
the first $d$ functions between the two codes, as plotted in Fig.~\ref{Fig.7}. 
As such, the DFT+$U$ implementations in ABACUS and OpenMX yield qualitatively similar
but quantitatively noticeably different results, as demonstrated in Fig.~\ref{Fig.1}.

\bibliographystyle{unsrt}
\bibliography{Reference_library}

\end{document}